\documentclass[pdftex,twocolumn,epjc3]{svjour3}          

\RequirePackage[T1]{fontenc}

\smartqed  

\RequirePackage{graphicx}
\RequirePackage{mathptmx}      
\RequirePackage{flushend}
\RequirePackage{ulem}
\RequirePackage[numbers,sort&compress]{natbib}
\RequirePackage[colorlinks,citecolor=blue,urlcolor=blue,linkcolor=blue]{hyperref}
\usepackage{amssymb}

\journalname{Eur. Phys. J. C}

\begin{document}

\title{Two-field cosmological phase transitions and gravitational waves in the singlet Majoron model}

\author{
Batool Imtiaz\thanksref{addr1, addr2, addr3}
\and
Youping Wan\thanksref{addr1, addr2, addr3}
\and
Yi-Fu Cai\thanksref{e3, addr1, addr2, addr3}
}

\thankstext{e3}{e-mail: yifucai@ustc.edu.cn}

\institute{Department of Astronomy, School of Physical Sciences, University of Science and Technology of China, Hefei, Anhui 230026, China \label{addr1}
\and
CAS Key Laboratory for Research in Galaxies and Cosmology, University of Science and Technology of China, Hefei, Anhui 230026, China \label{addr2}
\and
School of Astronomy and Space Science, University of Science and Technology of China, Hefei, Anhui 230026, China \label{addr3}
}

\date{Received: date / Accepted: date}

\maketitle

\begin{abstract}
In the singlet Majoron model, we study cosmological phase transitions (PTs) and their resulting gravitational waves (GWs), in the two-field phase space, without freezing any of the field directions. We first calculate the effective potential, at one loop and at finite temperature, of the Standard Model Higgs doublet together with one extra Higgs singlet. We make use of the public available Python package `CosmoTransitions' to simulate the two-dimensional (2D) cosmological PTs and evaluate the gravitational waves generated by first-order PTs. With the full 2D simulation, we are able not only to confirm the PTs' properties previously discussed in the literature, but also we find new patterns, such as strong first-order PTs tunneling from a vacuum located on one axis to another vacuum located on the second axis. The two-field phase space analysis presents a richer panel of cosmological PT patterns compared to analysis with a single-field approximation. The PTGW amplitudes turn out to be out of the reach for the space-borne gravitational wave interferometers such as LISA, DECIGO, BBO, TAIJI and TianQin when constraints from colliders physics are taken into account.
\end{abstract}

\section{Introduction}
\label{intro}
Strong first-order phase transitions (PTs) can be involved in a rich phenomenology for cosmology. In particular, they are conducive to successful baryogenesis \cite{Kuzmin:1985mm}---the mechanism which explains why there is more matter than anti-matter in our universe. On the other hand, the first discovery of gravitational waves (GWs), by the Advanced Laser Interferometer Gravitational Wave Observatory (aLIGO) \cite{Abbott:2016blz}, has pushed us to a new era of GW astronomy. We are now in the good position of studying GW generation in various physical processes, and searching them on multi-band frequencies. Today the stochastic GW background generated during first-order PTs \cite{Witten:1984rs, Hogan:1986qda, Kosowsky:1991ua, Kamionkowski:1993fg} has become more and more popular due to their rich phenomenology. Unfortunately, the electroweak phase transition (EWPT) in standard model (SM) turns out to be a crossover \cite{Kajantie:1993ag,Fodor:1994sj,Kajantie:1995kf}, which is too weak to get baryogenesis or PTGW. In order to realize strong first-order PTs and probe new physics, various extensions of the SM have been studied, such as models with extra scalar singlet(s) \cite{Anderson:1991zb, Espinosa:1993bs, Espinosa:2007qk, Profumo:2007wc, Espinosa:2011ax, Huang:2015bta}, with dimensional six effective operators \cite{Zhang:1992fs, Grojean:2004xa, Delaunay:2007wb, Huang:2015izx, Huang:2016odd, Cai:2017tmh}, 2HDMs \cite{Cline:1996mga, Dorsch:2013wja, Basler:2016obg, Dorsch:2017nza, Basler:2017uxn}, NMSSM \cite{Apreda:2001us, Huber:2007vva, Huber:2015znp}; see also other studies \cite{Espinosa:2008kw, Dev:2016feu, Schwaller:2015tja, Huang:2016cjm, Huang:2017laj, Chao:2017vrq, Addazi:2017oge, Chen:2017cyc, Cai:2018rzd, Li:2018ixg, Ezquiaga:2018btd, Hashino:2018wee}.

The singlet Majoron model is a simple extension of SM \cite{Chikashige:1980ui} (see also the famous GR model \cite{Gelmini:1980re}), which was originally proposed to solve the neutrino mass problem. In this model, an additional complex Higgs singlet is introduced, which breaks the global $U(1)_{B-L}$ symmetry with its VEV and generates mass for the right-handed (RH) neutrino. Recently, there has arisen strong interest in the possibility of first-order phase transitions \cite{Kondo:1991jz, Sei:1992np, Enqvist:1992va, Cline:2009sn}, GWs in radio astronomy \cite{Addazi:2017nmg} and dark matter physics \cite{Queiroz:2014yna}. In Refs. \cite{Kondo:1991jz, Sei:1992np} the authors claim that a flat direction exists on the surface of the effective Majoron potential, allowing one to reduce the two-field problem to a one-effective-field problem. Their analysis shows that strong first-order PTs may be realized. Another study \cite{Enqvist:1992va}, further confirmed PTs can typically proceed in two steps, i.e., a very weakly first-order transition from the $U(1)_{B-L}$ breaking followed by the EWPT. In this earlier work, the Yukawa couplings between the singlet Higgs field and the right-handed neutrinos was not considered. Later on, a very comprehensive work was completed \cite{Cline:2009sn} by a full numerical simulation to handle the two-field problem. Their work shows the existence of two step PTs---and once again the PT due to the $U(1)_{B-L}$ symmetry breaking is above the EWPT. They also find large values of the two Higgs interactive coupling while the Yukawa coupling between the Higgs singlet and the right-handed neutrinos are required in order to obtain strong first-order PTs.

In this paper, we conduct a full numerical simulation to analyze the two-field cosmological PTs and evaluate the resulting GWs, by using the public available Python package CosmoTransition \cite{Wainwright:2011kj}. Thanks to the full 2D simulation, we are able to not only confirm the PT patterns which have been found in \cite{Enqvist:1992va, Cline:2009sn}, but also find more new patterns, see Sect. \ref{CPTs} for more details. For each PT pattern we have found, we discuss the possibility to detect PTGWs with space-borne gravitational wave interferometers such as LISA \cite{Audley:2017drz}, DECIGO \cite{Seto:2001qf, Kawamura:2006up, Kawamura:2011zz, Kudoh:2005as}, BBO \cite{Corbin:2005ny}, TAIJI (ALIA descoped) \cite{Gong:2014mca} and TianQin \cite{Luo:2015ght, Hu:2018yqb}.

The paper is organized as follows: in Sect. \ref{model} we write down the singlet Majoron model and set up our notations; in Sect. \ref{CPTs} we calculate the effective potential of the two classic fields within the finite temperature field theory, and present patterns of cosmological PTs produced by our simulation that we classify; in Sect. \ref{PTGWs} we evaluate the GWs generated from first-order PTs and put them in the light of future space-borne GW detectors; finally, in Sect. \ref{conclusion} we summarize our main results and conclude.

\section{The singlet Majoron model}
\label{model}

The singlet Majoron model is among the simplest extensions of the Standard Model (SM) of particle physics. An $U(1)_{B-L}$ global symmetry, where the $B$ and $L$ stand for baryon and lepton numbers, is introduced in addition to the original gauge symmetry $SU_{c}(3) \times SU(2)_{L} \times U(1)_{Y}$. Right-handed neutrinos $\nu_{R}$, together with a complex singlet Higgs $\sigma$ become new members of the high energy physics particles zoo. The Lagrangian starts from
\begin{eqnarray}
\label{Lagrangian}
 \mathcal{L} = -f\bar{L}i\sigma_2 \Phi^{*}\nu_{R}-g\sigma \bar{\nu}_{R}\nu_{R}^{c}+h.c. ~,
\end{eqnarray}
in which $\Phi$ and $\bar{L}$ are the SM Higgs and Light-handed(LH) fermions doublets, and $f$ and $g$ are Yukawa coupling constants.

The potential for the doublet and singlet Higgs at tree level can be written as
\begin{eqnarray}
\label{potential-tree}
 V(\sigma,\Phi) = &\lambda_s |\sigma|^4 +\mu_s^2|\sigma|^2 +\lambda_h |\Phi|^4 +\mu_h^2|\Phi|^2 \\ \nonumber
 & +\lambda_{sh} |\sigma|^2|\Phi|^2 ~.
\end{eqnarray}
In order to ensure today's vacuum $\left(v_{BL},~v_{ew}=246~{\rm GeV}\right)$ is realized, we will choose $\mu_s^2 = -\lambda_s v_{BL}^2 -(1/2)\lambda_{sh}v^2_{ew}$, $\mu_h^2 = -\lambda_h v_{ew}^2 -(1/2)\lambda_{sh}v^2_{BL}$. To be clearer, let us expand the Higgs Bosons into their real components
\begin{eqnarray}
\label{UG-01}
 \sigma &&= \frac{1}{\sqrt{2}}\left(v_{BL}+\rho+i\chi\right) ~,~~ \\
 \Phi^{T} &&= \frac{1}{\sqrt{2}}\left(G_1+iG_2,~v_{ew}+H+iG_3\right)^{T} ~.
\end{eqnarray}
Just as a Dirac mass with $m={f v_{ew}}/(2\sqrt{2})$ is obtained after the electroweak symmetry breaking, a Majorana mass with $M={gv_{BL}}/{\sqrt{2}}$  is generated after the breaking of the $U(1)_{B-L}$ symmetry. If the Majorana mass is much larger than the Dirac one, we may redefine heavy and light neutrinos as $N = \nu_{R} +\nu_{R}^{c} +{m}/{M}\left(\nu_{L} +\nu_{L}^{c}\right)$, and $\nu = \nu_{L} +\nu_{L}^{c} -{m}/{M} \left(\nu_{R}^{c} +\nu_{R} \right)$.
Then the masses of light neutrinos, $m_{\nu}\simeq {m^{2}}/{M}$, are highly suppressed by the heavy neutrinos mass $m_{N}=M$. This is perhaps the simplest way to understand why neutrinos' observational masses are so small. However, this model cannot be the most satisfactory one when compared with the experimental data, from light neutrinos $m_{\nu}=(\sqrt{2}/8)(f^2/g)(v^2_{ew}/v_{BL})$; it would lead to an unnaturally small value for the Yukawa coupling $f<3\times 10^{-7}\sqrt{g}(v_{BL}/{\rm GeV})^{1/2}$ (by taking $m_{\nu}<2~{\rm eV}$). Therefore, in the present article we simply take it as a benchmark model in order to show in detail how multi-step phase transitions happen when there are more than one degrees of freedom being involved. The issue of unnatural smallness can be cured in some advanced models, namely, we refer to Refs. \cite {Berezhiani:1992cd, Berezhiani:2015afa} for the associated analyses.

A massless Goldstone boson $\chi$ appears after the $U(1)_{B-L}$ global symmetry is broken, this is the Majoron field discussed in this paper. Although, with the above potential, the Majoron is massless, a small mass term can be generated after one has considered effective operators with higher dimensions. Following the arguments of Ref. \cite{Akhmedov:1992hi}, we adopt a constraint
\begin{eqnarray}
 v_{BL}\lesssim\left(\frac{m_{\nu}}{25~{\rm eV}}\right)^{4/7}\times 10~{\rm TeV} ~.
\end{eqnarray}
Since the neutrino mass upper bound is smaller than $m_{\nu}\sim 2~{\rm eV}$ \cite{Patrignani:2016xqp}, it yields
\begin{eqnarray}
\label{constraint-on-vBL}
 v_{BL}\lesssim 1.6~{\rm TeV} ~.
\end{eqnarray}
In \cite{Fukugita:1990gb, Cline:1993ht}, it is pointed out that a $v_{BL}$ with a value larger than $v_{ew}$ can be dangerous. However, in our numerical simulation we will explore both scenarios with $v_{BL}$ larger or smaller than $v_{ew}$, in order to obtain a global outlook on the parameter space which can lead to strong first-order PTs and GWs. Remarkably, we find $v_{BL}>v_{ew}$ is also astronomically uninteresting according to the results in Sect. \ref{PTGWs}.

\section{Patterns of cosmological phase transitions in the singlet Majoron model}
\label{CPTs}

Consider
\begin{eqnarray}
\label{UG-02}
 \sigma&&=\frac{1}{\sqrt{2}}\left(s+\rho+i\chi\right) ~,~~ \\
 \Phi^{T}&&=\frac{1}{\sqrt{2}}\left(G_1+iG_2,~h+H+iG_3\right)^{T} ~.
\end{eqnarray}
In vacuum at the present scale one gets back Eq. (\ref{UG-01}). At tree level the potential can be rewritten as
\begin{eqnarray}
\label{tree}
 V^{\rm{tree}}(s, h) = &\frac{1}{4}\lambda_s \left(s^2-v_{BL}^2\right)^2 +\frac{1}{4}\lambda_h \left(h^2-v_{ew}^2\right)^2 \\ \nonumber
 & +\frac{1}{4}\lambda_{sh} \left(s^2-v_{BL}^2\right) \left(h^2-v_{ew}^2\right) ~,
\end{eqnarray}
whose global minimum clearly shows today's vacuum. Then consider the one loop correction at zero temperature with $\overline{\rm MS}$ renormalization
\begin{eqnarray}
 \Delta V_1^{T=0}(s,h)=\frac{1}{64\pi^2}\sum_i n_i m^4_i(s,h) \Big[ \log\frac{m^2_i(s,h)}{Q^2} -c_i \Big] ~,
\end{eqnarray}
where $(c_i, c_{W},c_{Z})=(3/2,5/6,5/6)$ and \\
$(n_H, n_G, n_{\rho}, n_{\chi}, n_{W}, n_{Z}, n_{\nu_R}, n_t) = (1,3,1,1,6,3,-6,-12)$. The mass spectrum can be found in Appendix \ref{appA}. We also introduce a counter term $\Delta V_{ct}^{\rm{T=0}}(s,h)=A h^2$ and use the following renormalization conditions:
\begin{eqnarray}
\label{RenormalizationConditions}
 \frac{\partial}{\partial s} \left(V^{\rm{tree}}+\Delta V_1^{T=0}+\Delta V_{ct}^{\rm{T=0}}\right) \Big{|}_{(v_{BL}, v_{ew})} =0 ~,\\
 \frac{\partial}{\partial h} \left(V^{\rm{tree}}+\Delta V_1^{T=0}+\Delta V_{ct}^{\rm{T=0}}\right) \Big{|}_{(v_{BL}, v_{ew})} =0 ~,
\end{eqnarray}
to make sure the tree level vacuum at zero temperature is not shifted. We then solve and substitute the energy scale $Q$ and $A$; see Appendix \ref{appA}. The temperature correction reads
\begin{eqnarray}
\label{TemperatureCorrectionPotential}
 \Delta V^{T\neq0}(s, h, T) = &\sum_F n_F \frac{T^4}{2\pi^2} J_F \left[ \frac{m_F^2(s,h)}{T^2} \right] \\ \nonumber
 &+\sum_B n_B \frac{T^4}{2\pi^2}J_B \left[ \frac{m_B^2(s,h)}{T^2} \right] ~,
\end{eqnarray}
with $J_i\left[\frac{m_i^2}{T^2}\right]=\int^{\infty}_{0}dx x^2 \log\left[ 1\pm\exp \left( -\sqrt{x^2+\frac{m_i^2}{T^2}} \right) \right]$; the plus sign is for fermions and the minus sign for bosons. In order to include the {\it{ring}} (or {\it{daisy}}) contribution, we make the following replacements \cite{Parwani:1991gq, Espinosa:1993bs, Cline:1996mga} for the bosonic masses:
\begin{eqnarray}
\label{Replacement}
 m^2_{B}(s,h)\rightarrow {\cal M}^2_B(s,h,T)=m^2_{B}(s,h)+\Pi_B(T) ~.
\end{eqnarray}
The expressions for the self-energies and the thermal mass (Debye mass) for the longitudinal component of $Z$ boson can be found in Appendix \ref{appA}.
We substitute the Debye mass terms into Eq. (\ref{TemperatureCorrectionPotential}). By doing so, the replacement Eq. (\ref{Replacement}) is applied in the effective potential (see the analyses in \cite{Cline:1996mga, Parwani:1991gq}), not only for the cubic term (see the analyses in \cite{Espinosa:1993bs, Cline:1996mga, Arnold:1992rz}). The two different prescriptions can be understood as the theoretical uncertainty in one-loop perturbation theory, which can be safely handled.
Finally, the total effective potential of classic fields is given by
\begin{eqnarray}
\label{effective potential}
 V_{\rm{eff}}(s,h,T) &=& V^{\rm{tree}}(s,h)+\Delta V_1^{T=0}(s,h) \\ \nonumber
  && +\Delta V_{ct}^{\rm{T=0}}(s,h) +\Delta V^{T\neq0}(s,h,T) ~.
\end{eqnarray}

With the above effective potential, the bubble profiles can be found by solving the bounce equation
\begin{eqnarray}
 \frac{d^2 \vec{\phi}}{d r^2} +\frac{\alpha}{r}\frac{d\vec{\phi}}{dr} = \nabla V_{\rm{eff}}(\vec{\phi}) ~, \\ \nonumber
 \vec{\phi}(r \rightarrow \infty) = \vec{\phi}_F ~,~~ \frac{d\vec{\phi}}{dr}\Big{|}_{r=0}=0 ~,
\end{eqnarray}
where $r^2=|\vec{x}|^2$, $\alpha=2$ at finite temperature, and $r^2=|\vec{x}|^2-t^2$, $\alpha=3$ at zero temperature. For single field problems, they can be solved by the so-called `overshooting-undershooting' method \cite{Coleman:1977py, Callan:1977pt}; however, the case of multiple fields becomes much more complicated. To overcome this situation, a powerful method dubbed `path deformation' was developed in Ref. \cite{Wainwright:2011kj}. In our analysis, we make use of the public available package CosmoTransition \cite{Wainwright:2011kj} to study the PTs in the case of multiple fields.

The strength of the PTs can be illustrated by the parameter $\alpha$, which is the ratio between the latent heat and the radiation energy,
\begin{eqnarray}
 \alpha \equiv \frac{\rho_{\rm vac}}{\rho_{r}}\Big{|}_{T=T_*} ~,
\end{eqnarray}
where the latent heat is evaluated using \\
$\rho_{\rm vac} = \big[ T(d\Delta V_{\rm eff}/d T) -\Delta V_{\rm eff} \big]{|}_{T=T_*}$, $\Delta V_{\rm eff}$ is the potential difference between the true and false vacua; the radiation energy density is $\rho_r =g_* \pi^2 T^4_* /30$ with $g_*$ being the relativistic degrees of freedom. The time scale of the PT is the inverse of the parameter $\beta$,
\begin{eqnarray}
\beta = -\frac{d(S_3/T)}{dt}\Big{|}_{t=t_*} \simeq \frac{1}{\Gamma}\frac{d\Gamma}{dt}\Big{|}_{t=t_*} ~,
\end{eqnarray}
where the Euclid action is \\
$S_3 = \int d^3\vec{x} \big[ (1/2)(\nabla s)^2+(1/2)(\nabla h)^2+V_{\rm eff}(s,h,T) \big]$, and the bubble nucleation rate defined by $\Gamma=\Gamma_0 \exp{(-S_3/T)}$. In the actual calculation, a renormalization of $\beta$ is quite useful:
\begin{eqnarray}
 \tilde{\beta}\equiv\frac{\beta}{H_*} = T_* \frac{d(S_3/T)}{dT}\Big{|}_{T=T_*} ~.
\end{eqnarray}
By definition, a larger value of $\alpha$ means a stronger PT, and a larger $\beta$ means a faster PT. The PT temperature $T_*$ is estimated when the bubble nucleation probability in units of time and units of volume $\simeq 1$,
\begin{eqnarray}
 & 1\simeq\int^{t_*}_{0}\frac{\Gamma dt}{H^3}= \\ \nonumber
 &\int^{\infty}_{T_*}\frac{dT}{T} \left( \frac{90}{8\pi^3 g_{\rm eff}} \right)^2 \left( \frac{M_{pl}}{T}\right)^4 \exp{(-S_3/T)} ~.
\end{eqnarray}
This gives us a simple criterion to estimate $T_*$ as follows,
\begin{eqnarray}
\label{criterion of Tstar}
 \frac{S_3}{T_*} \simeq \ln{ \left[ \frac{1}{4} \left( \frac{90}{8\pi^3 g_{\rm eff}} \right)^2 \right]} +4\ln{ \left[ \frac{M_{pl}}{T_*} \right] } ~.
\end{eqnarray}
Note that on when we deriving this criterion, we have assumed that the energy density of the universe is dominated by radiation, so it is not applicable when $\alpha$ is very large---which means the radiation dominated condition is spoiled, e.g., when bubbles show runaway in the vacuum \cite{Caprini:2015zlo}.

\begin{figure*}
\resizebox{0.33\textwidth}{!}{%
  \includegraphics{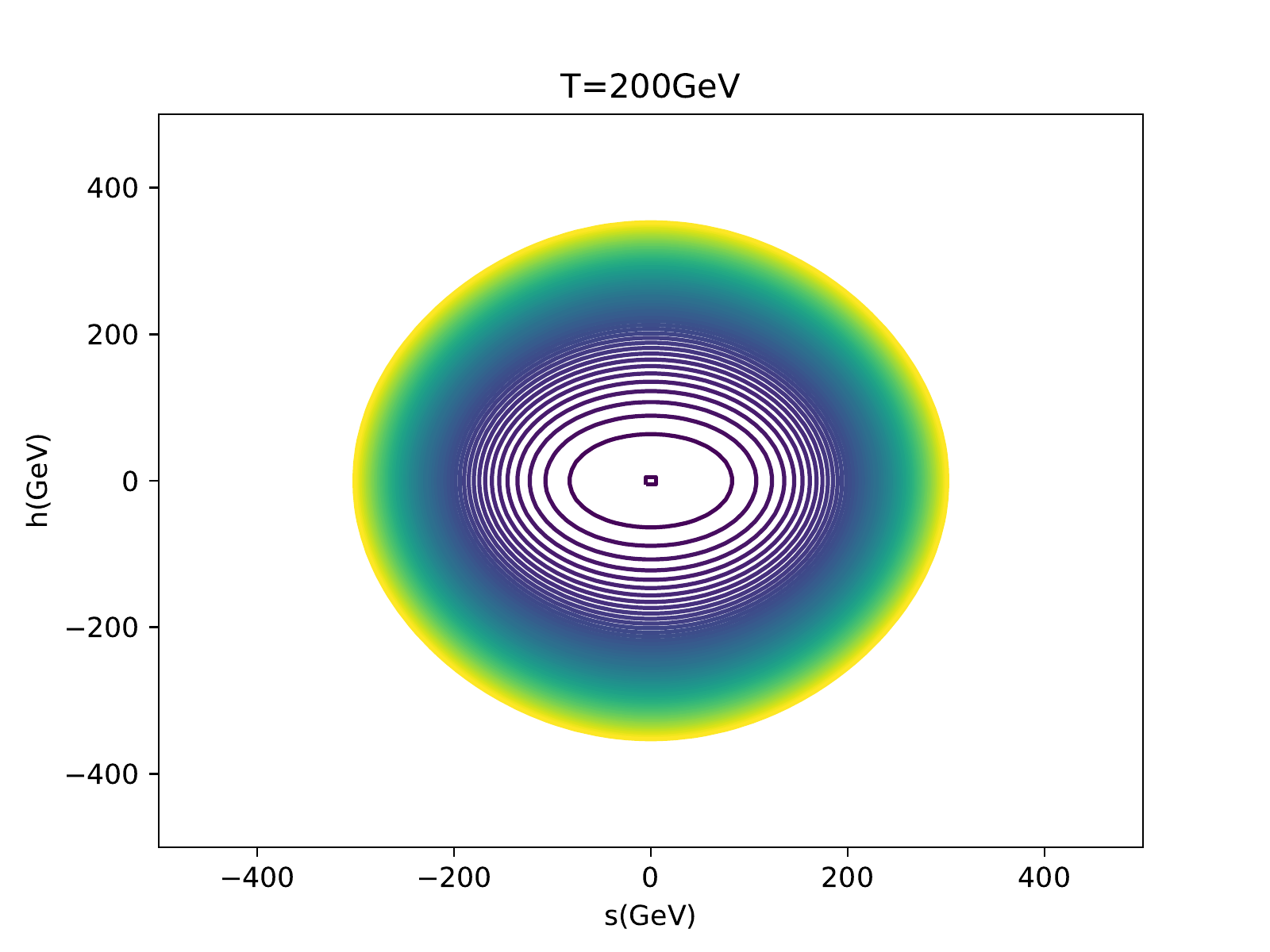}
}
\resizebox{0.33\textwidth}{!}{%
  \includegraphics{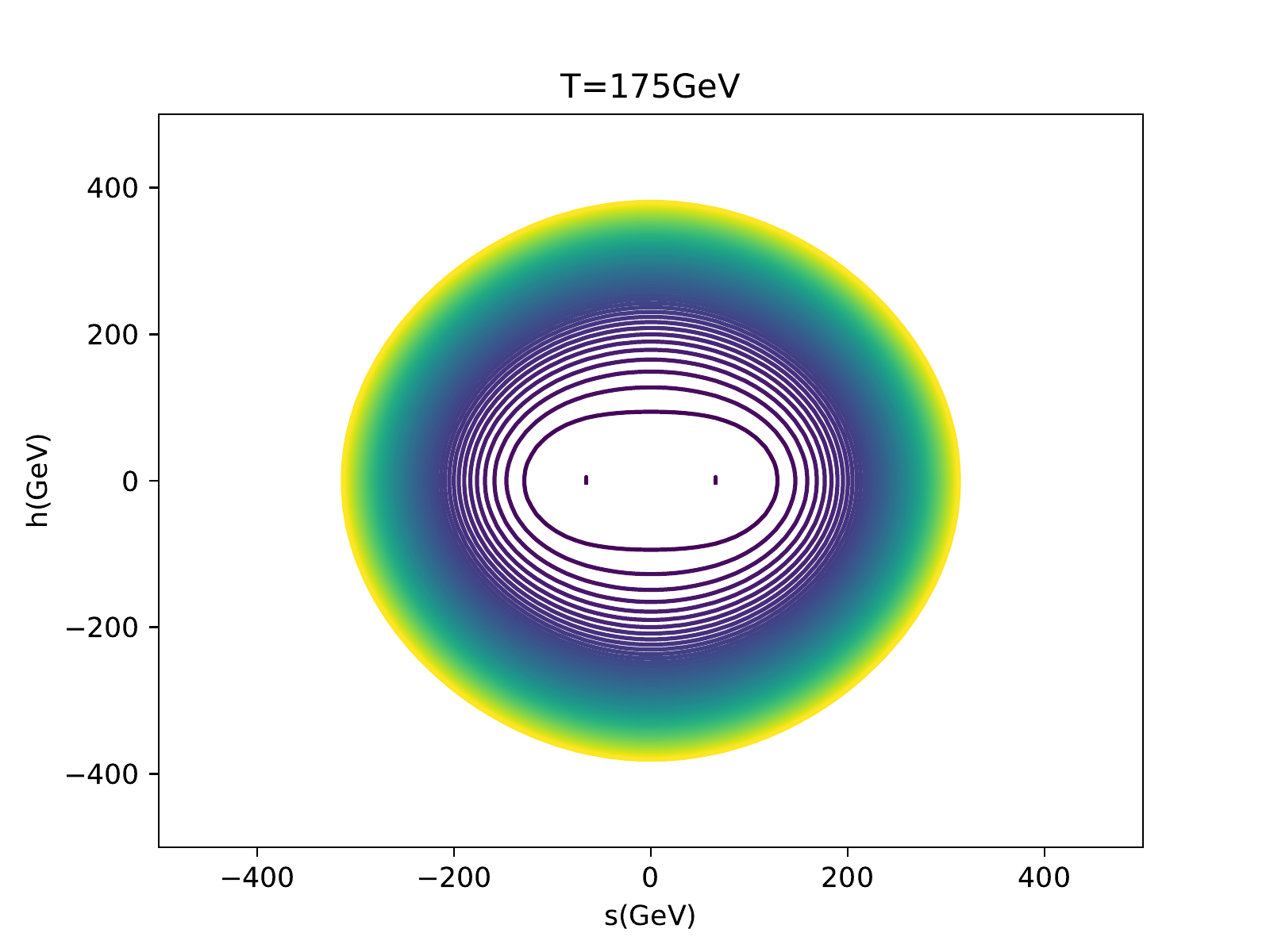}
}
\resizebox{0.33\textwidth}{!}{%
  \includegraphics{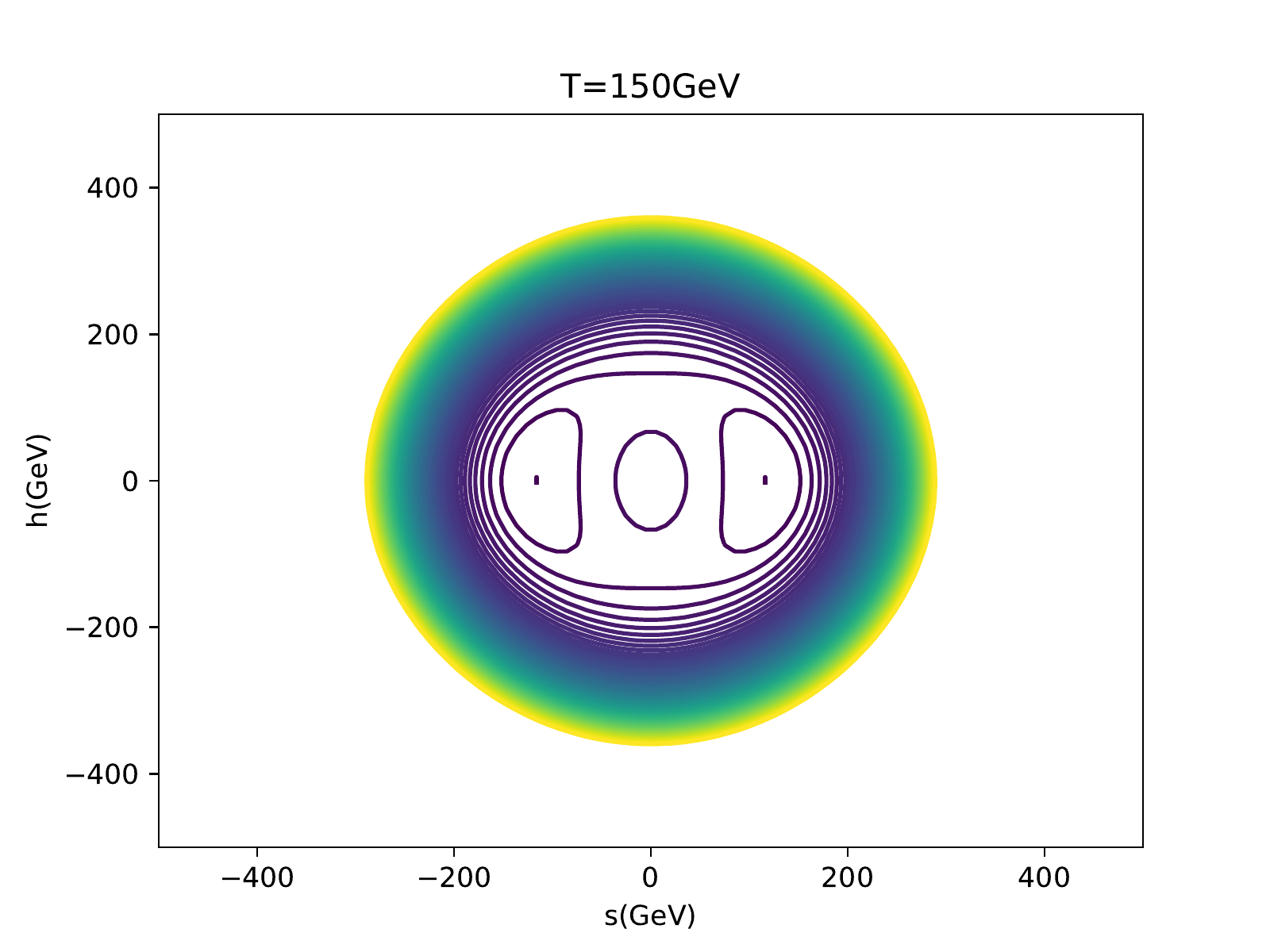}
}
\resizebox{0.33\textwidth}{!}{%
  \includegraphics{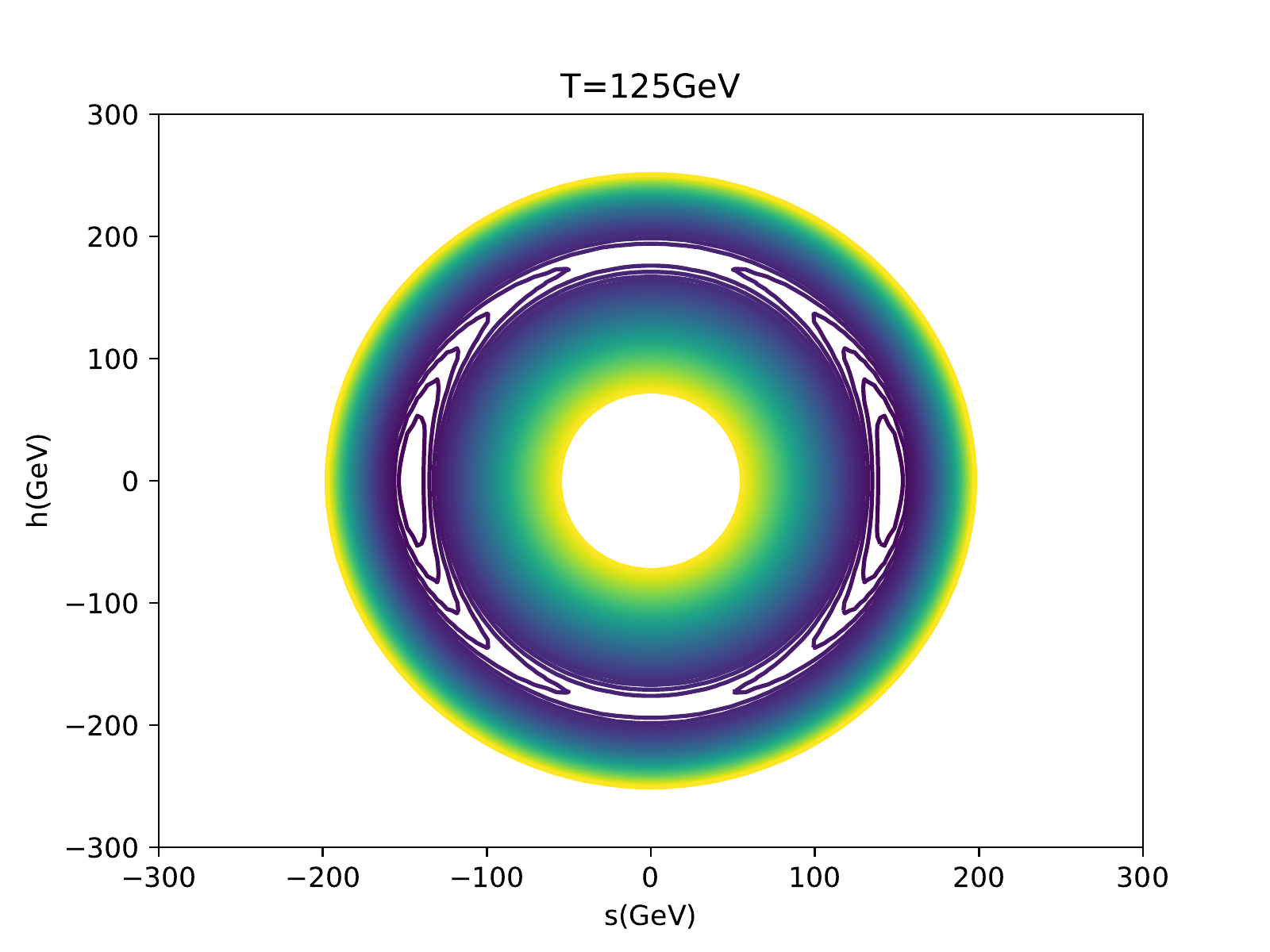}
}
\resizebox{0.33\textwidth}{!}{%
  \includegraphics{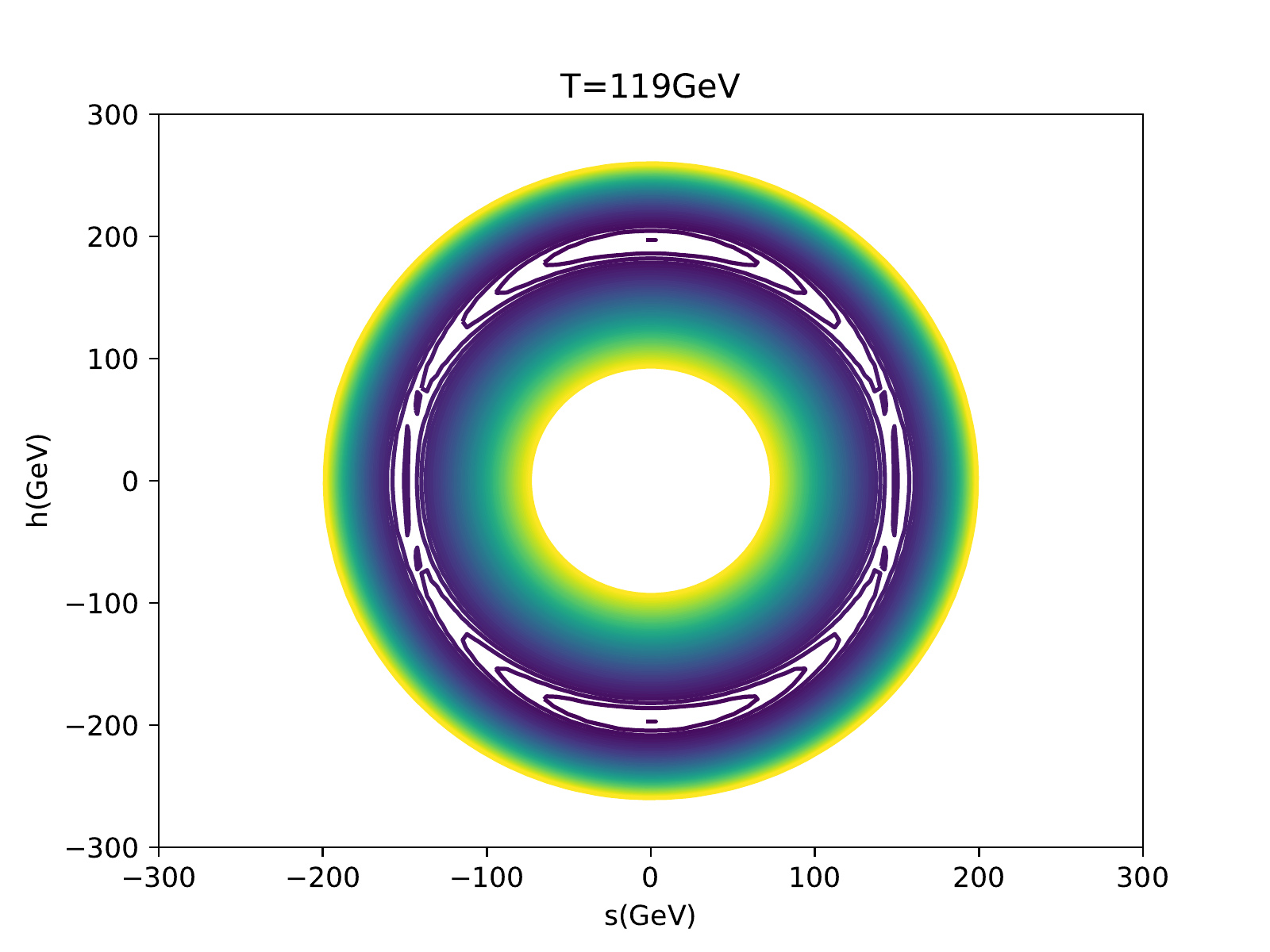}
}
\resizebox{0.33\textwidth}{!}{%
  \includegraphics{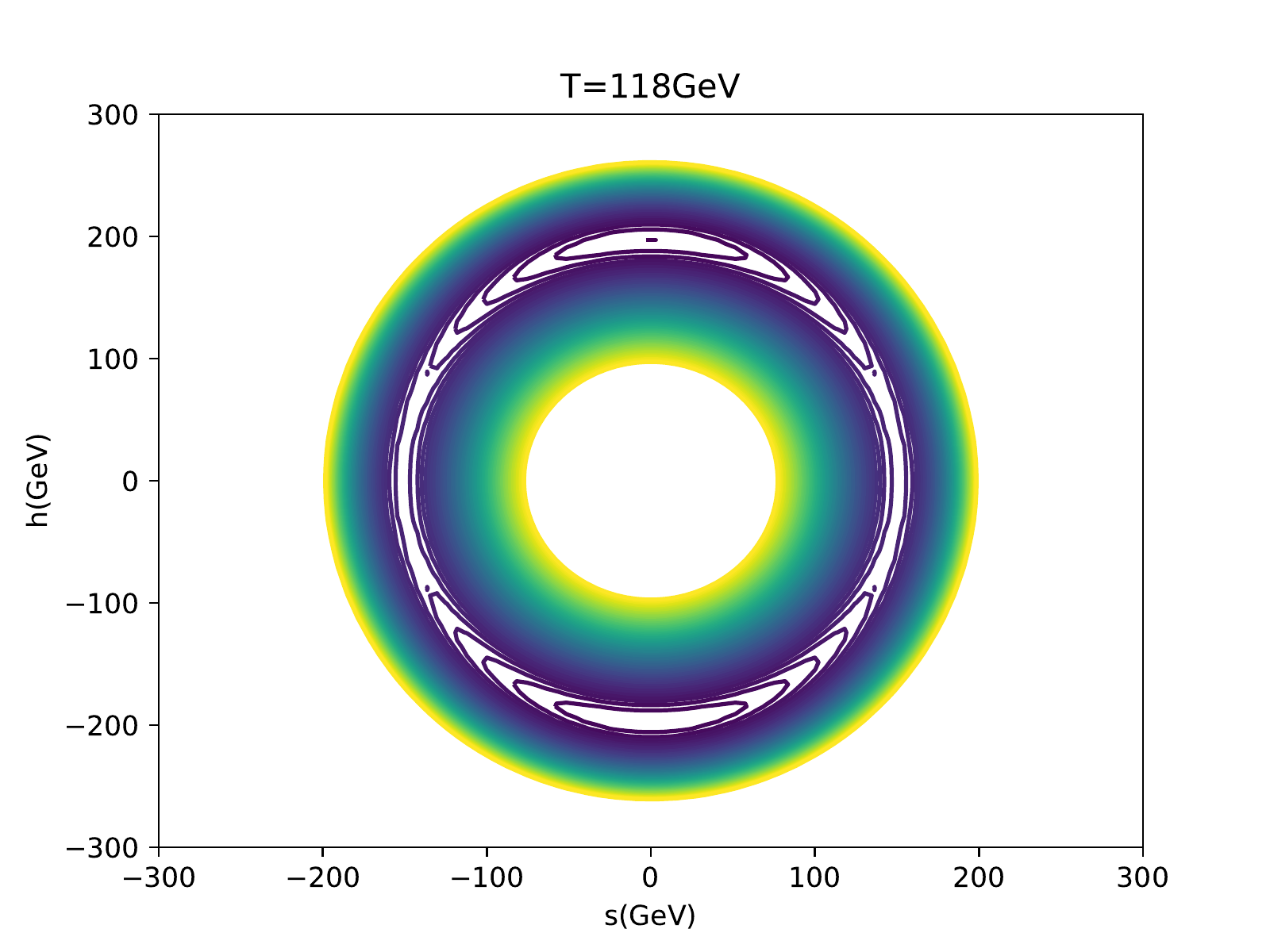}
}
\resizebox{0.33\textwidth}{!}{%
  \includegraphics{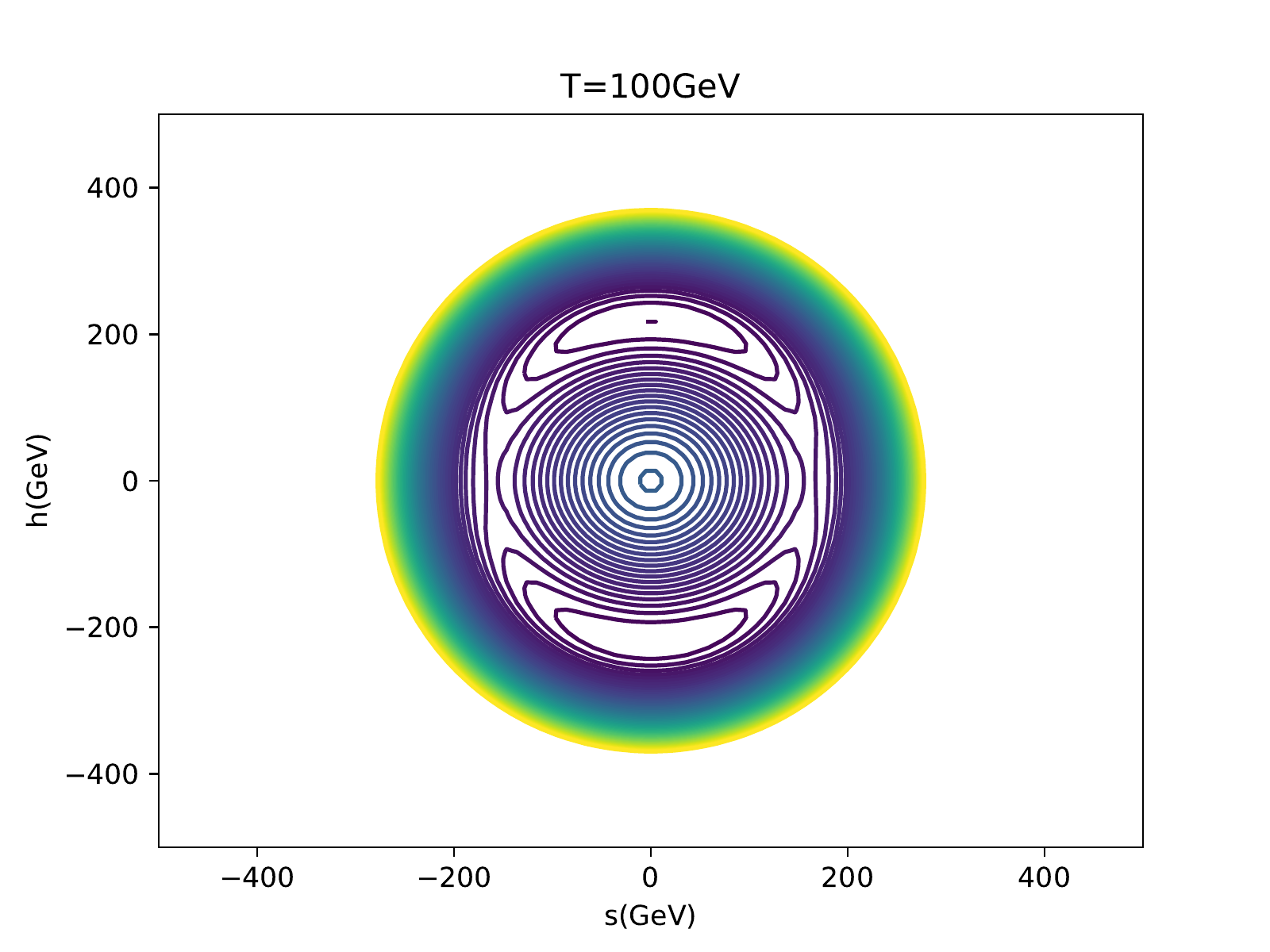}
}
\resizebox{0.33\textwidth}{!}{%
  \includegraphics{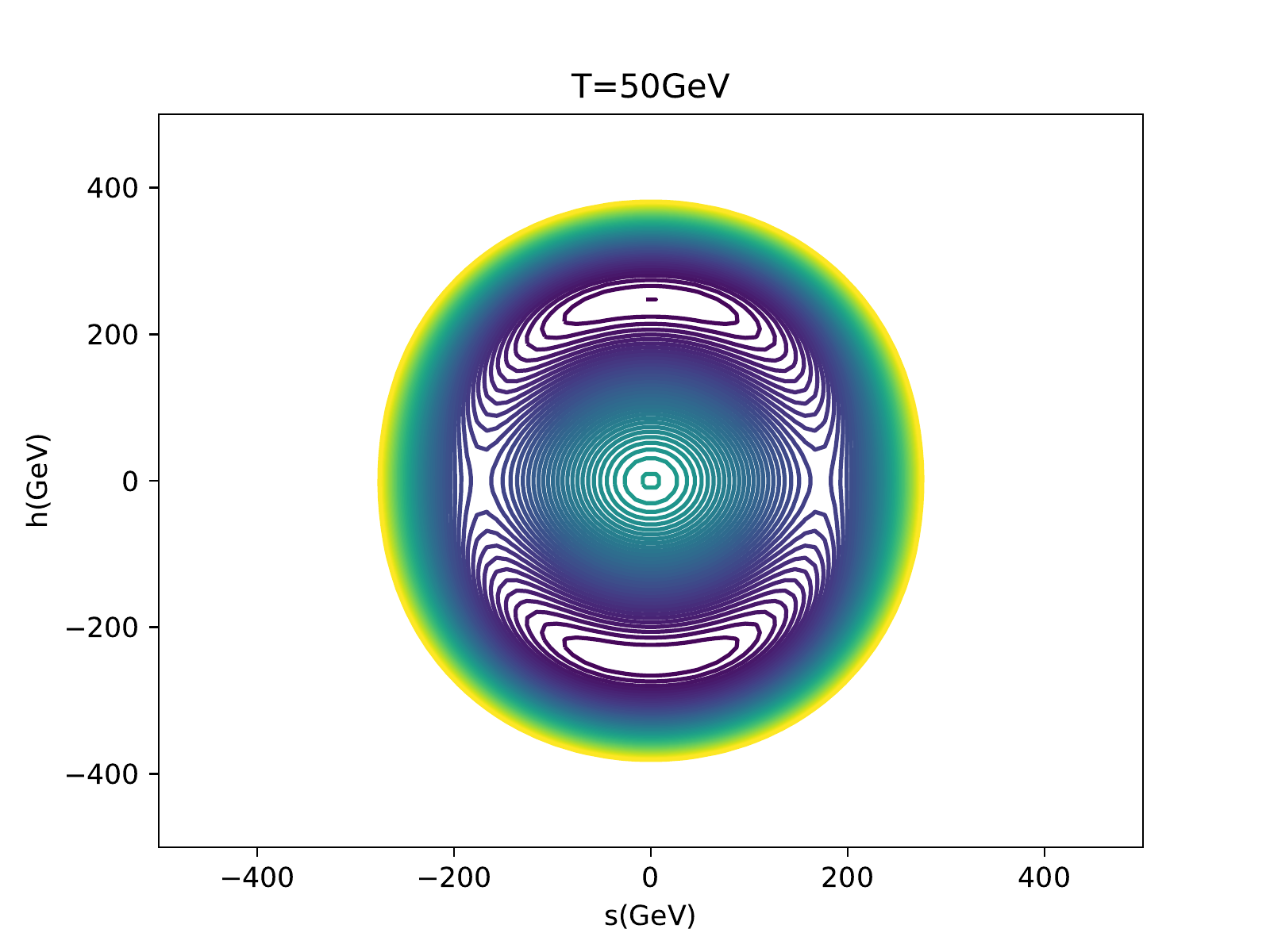}
}
\resizebox{0.33\textwidth}{!}{%
  \includegraphics{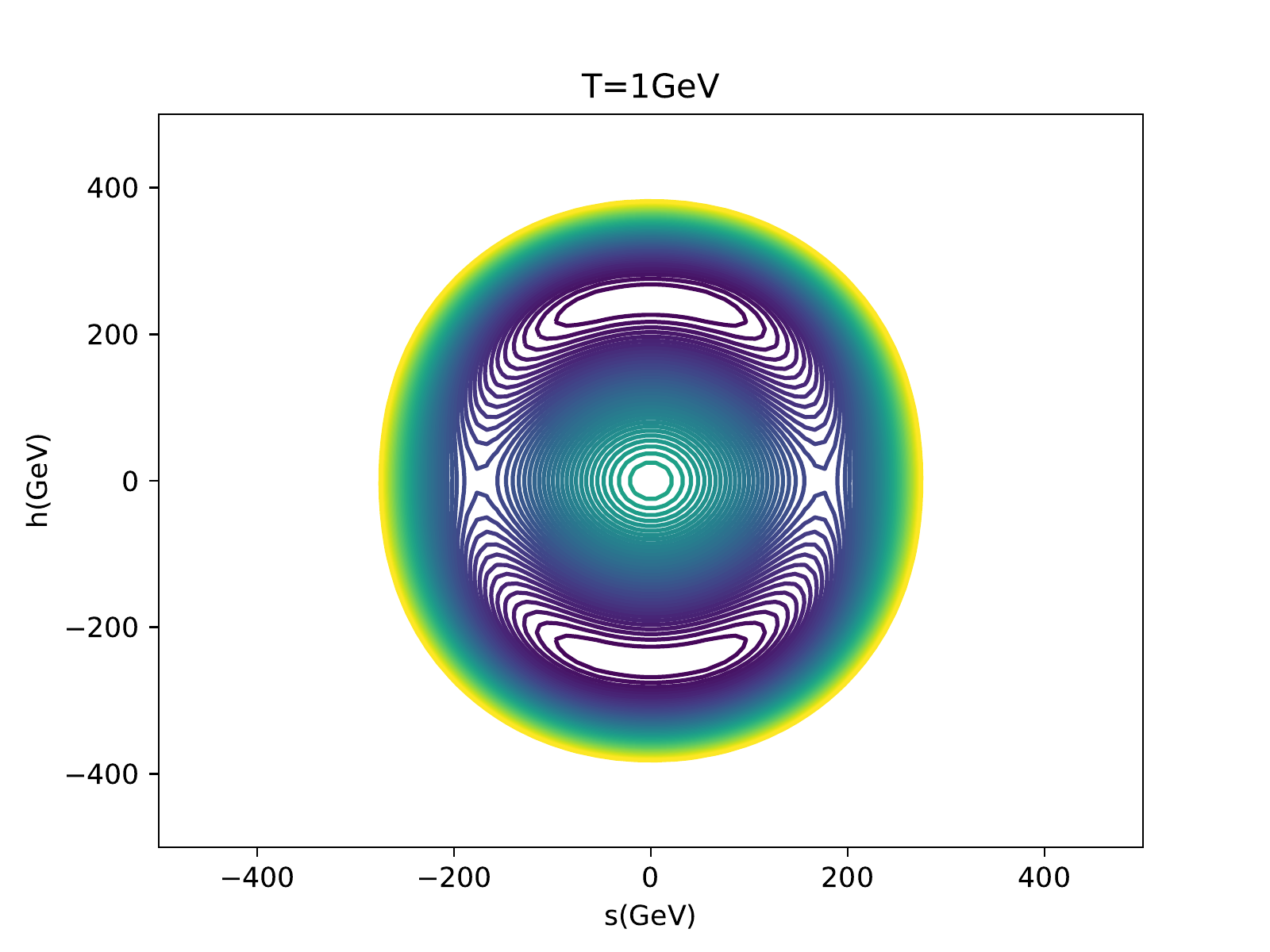}
}
\caption{An example of a higher order $s$-direction PT followed by a hybrid first-order PT. This is given by the parameters $v_{BL}=25~{\rm GeV}$, $\lambda_s=0.35$, $\lambda_{h}=0.126$, $\lambda_{sh}=0.388$, $g=1.5$. Here the higher-order PT happens around $T\simeq 182~{\rm GeV}$, the hybrid PT happens at $T\simeq 118.89~{\rm GeV}$.}
\label{Hybrid}
\end{figure*}

\begin{figure*}
\resizebox{0.33\textwidth}{!}{%
  \includegraphics{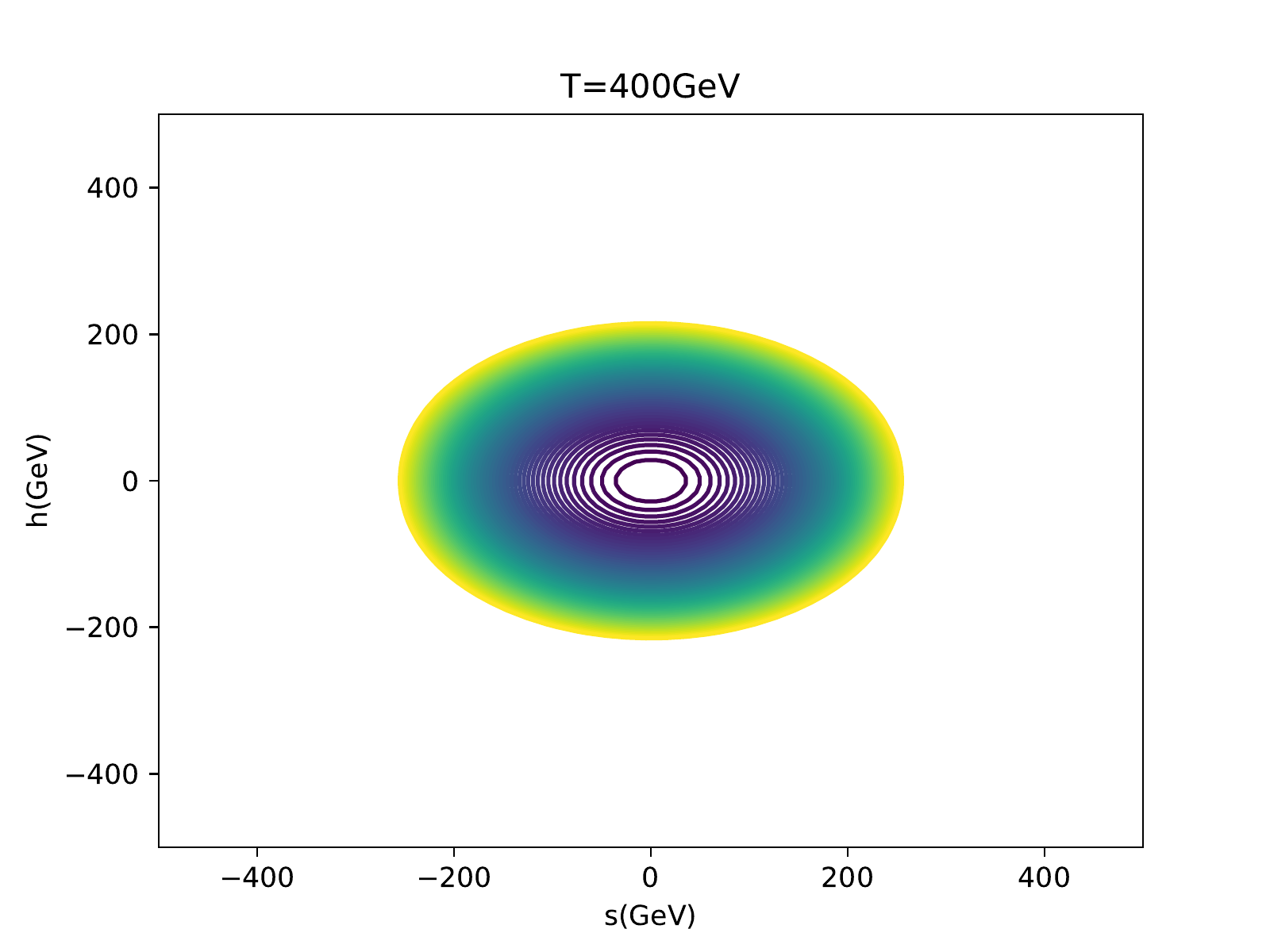}
}
\resizebox{0.33\textwidth}{!}{%
  \includegraphics{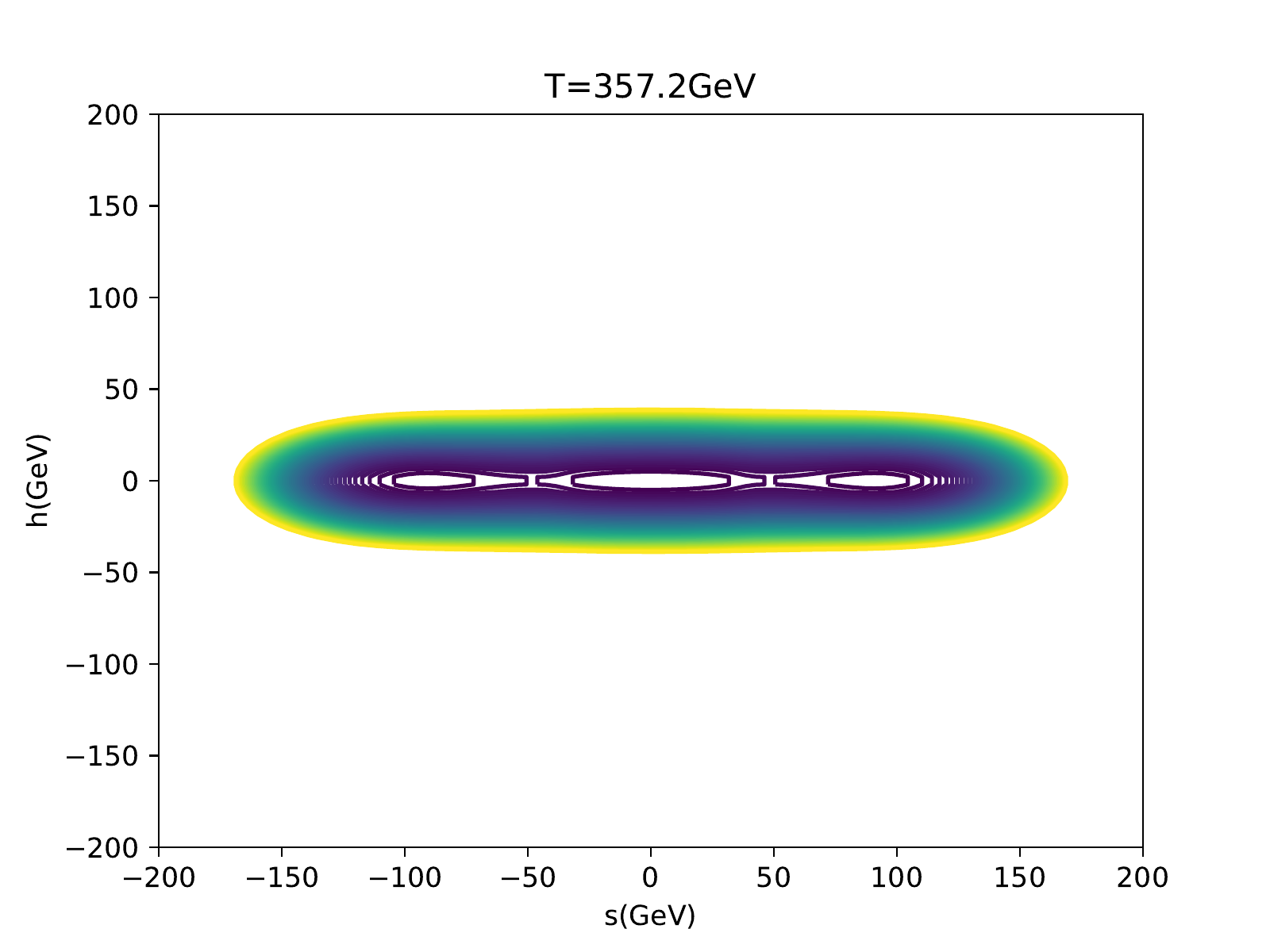}
}
\resizebox{0.33\textwidth}{!}{%
  \includegraphics{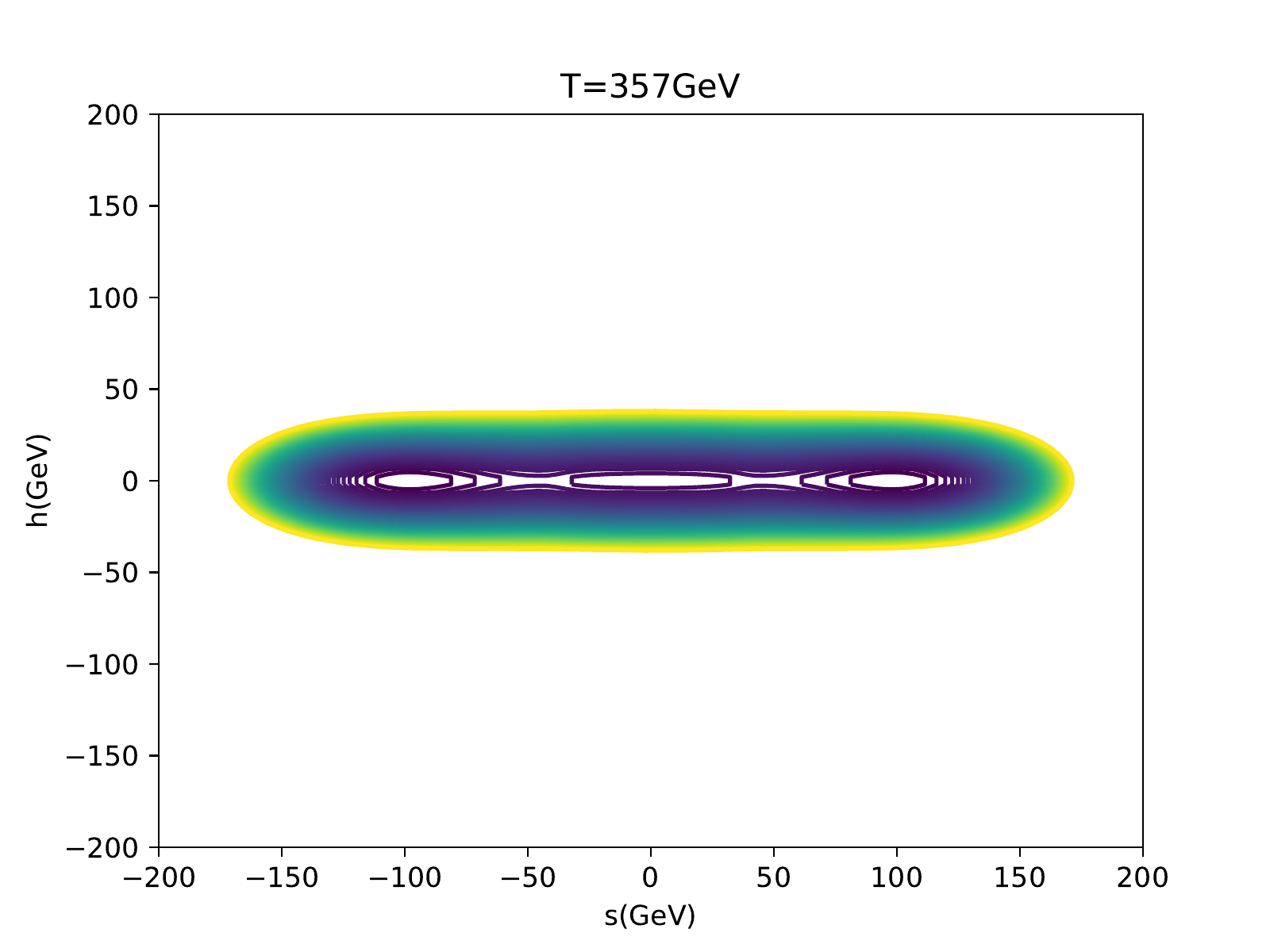}
}
\resizebox{0.33\textwidth}{!}{%
  \includegraphics{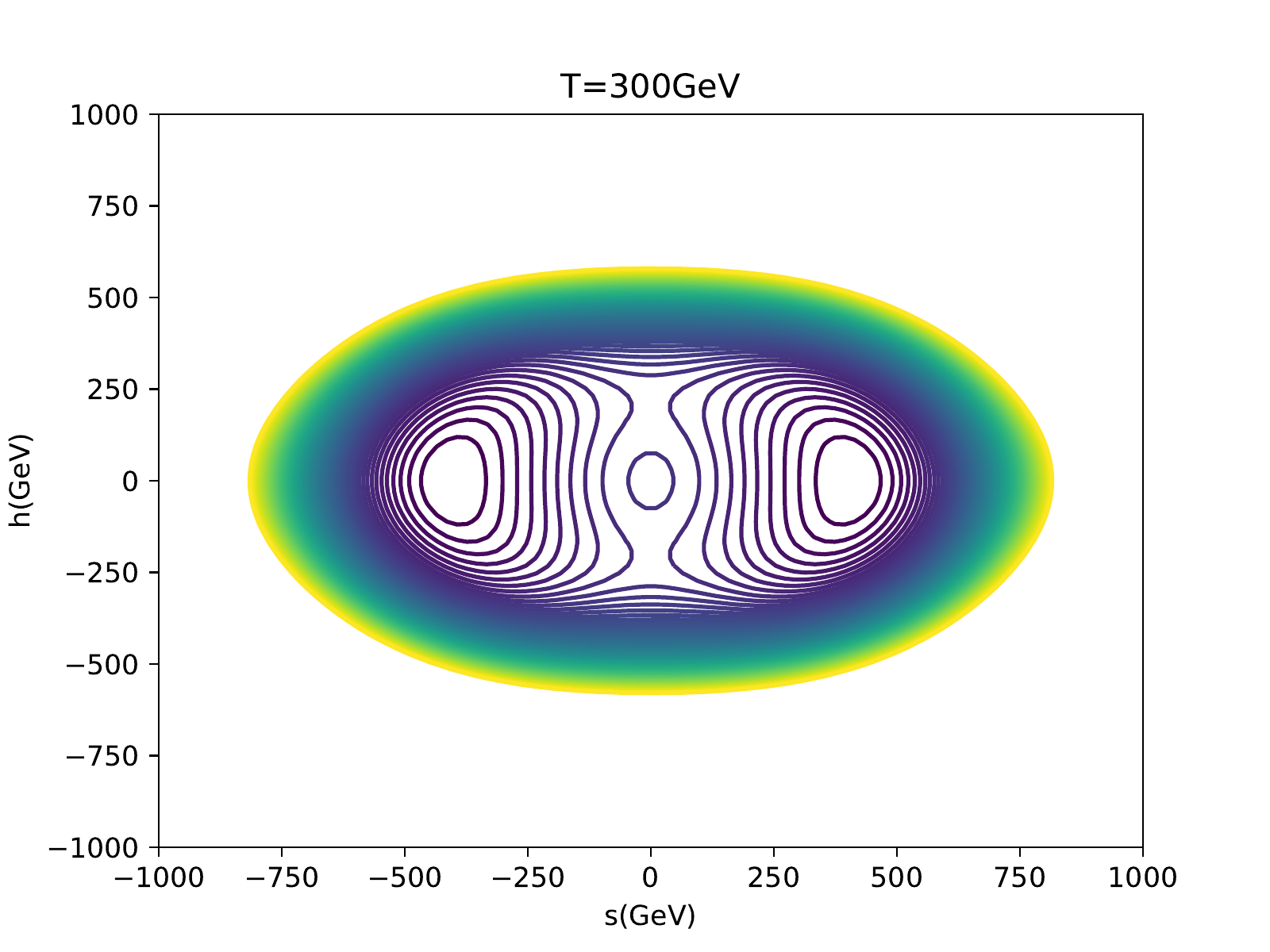}
}
\resizebox{0.33\textwidth}{!}{%
  \includegraphics{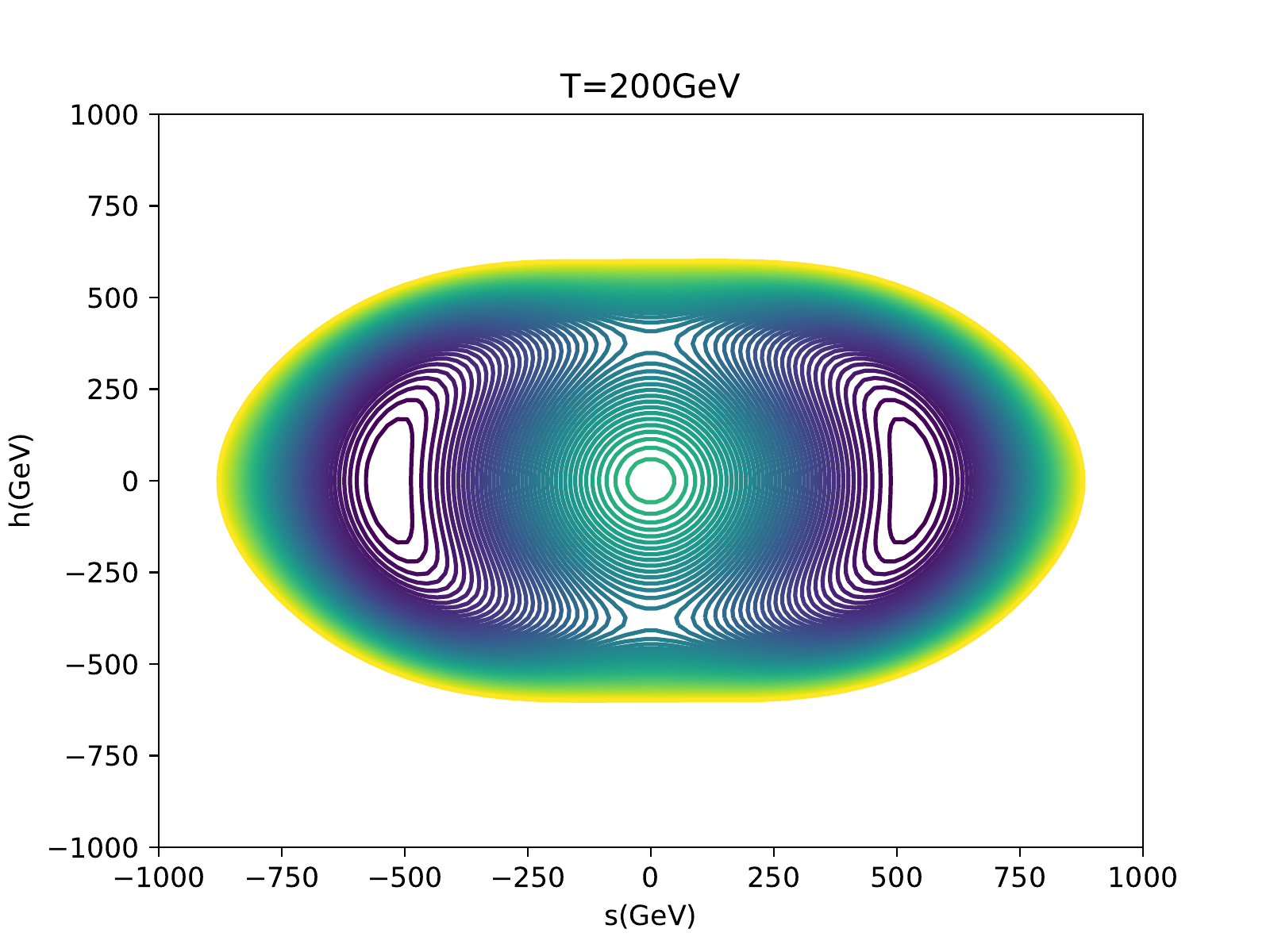}
}
\resizebox{0.33\textwidth}{!}{%
  \includegraphics{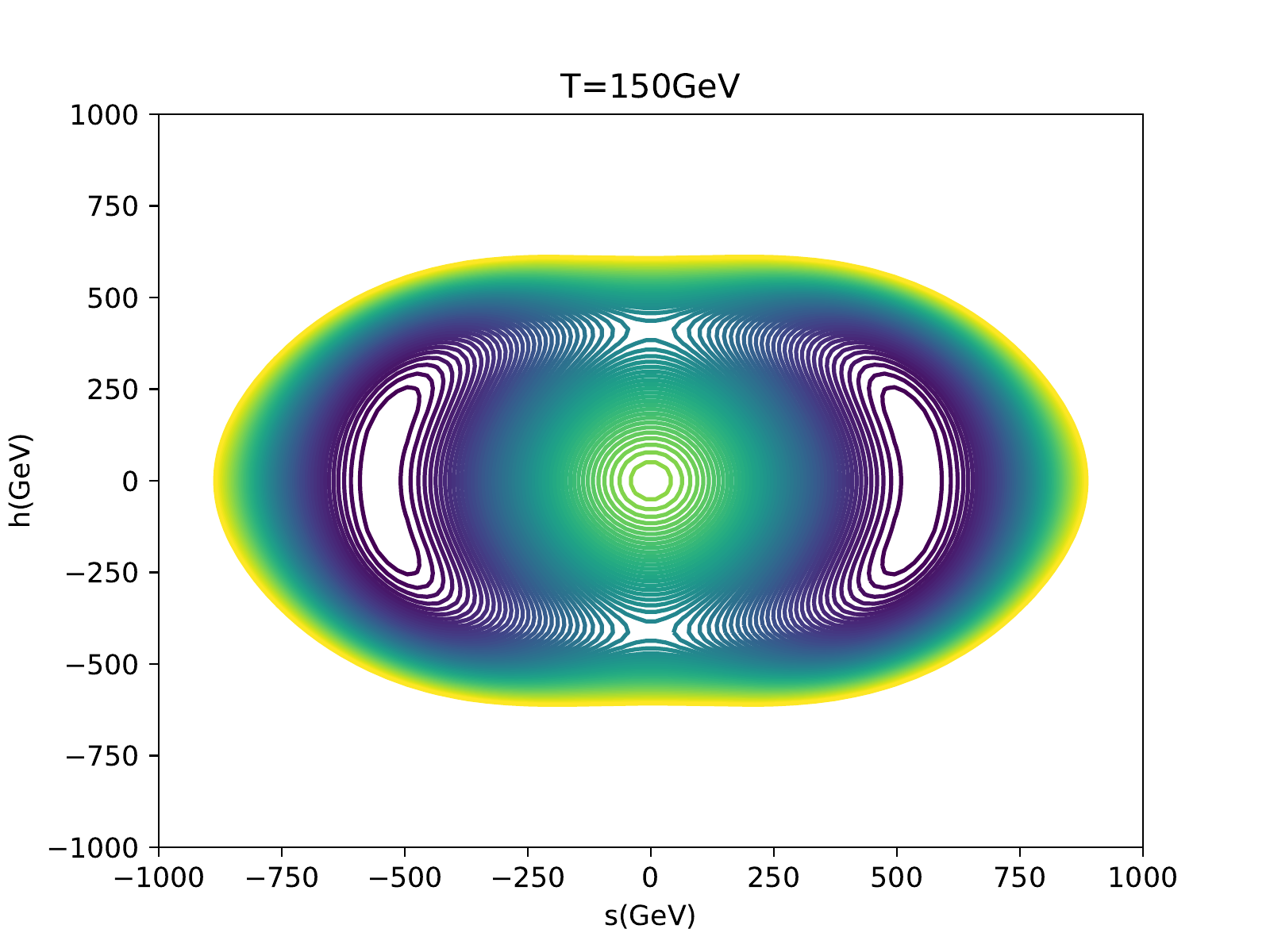}
}
\resizebox{0.33\textwidth}{!}{%
  \includegraphics{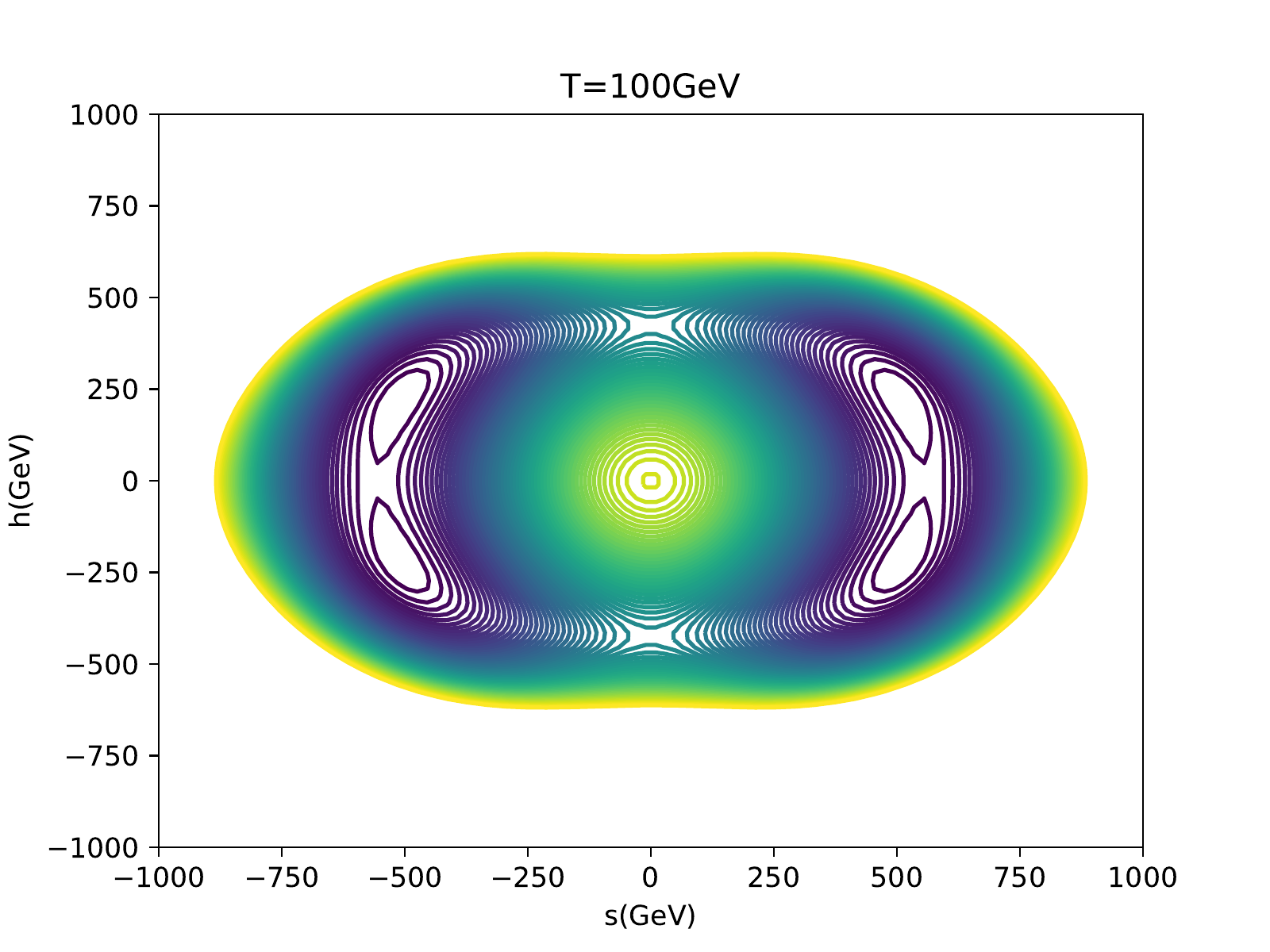}
}
\resizebox{0.33\textwidth}{!}{%
  \includegraphics{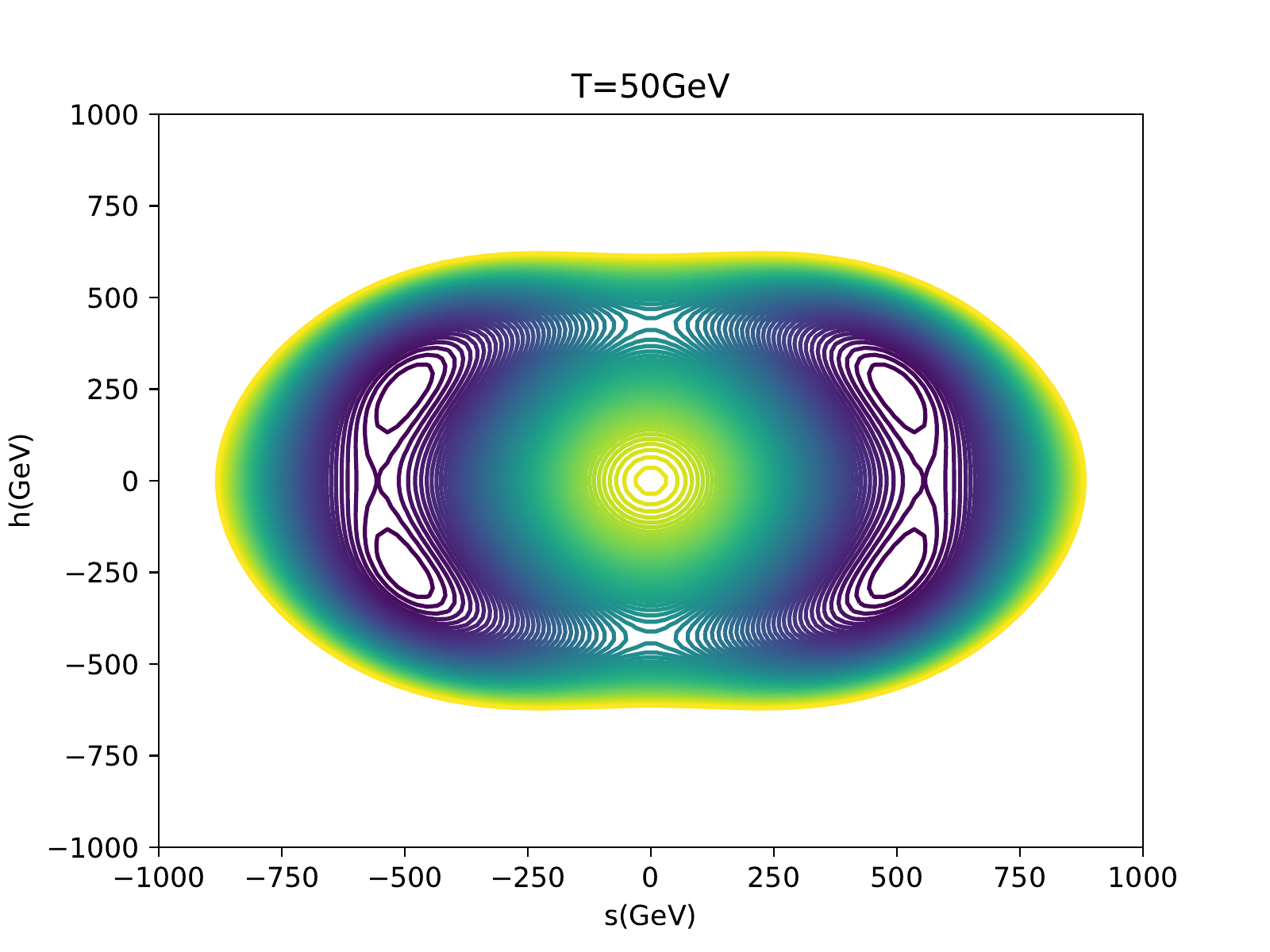}
}
\resizebox{0.33\textwidth}{!}{%
  \includegraphics{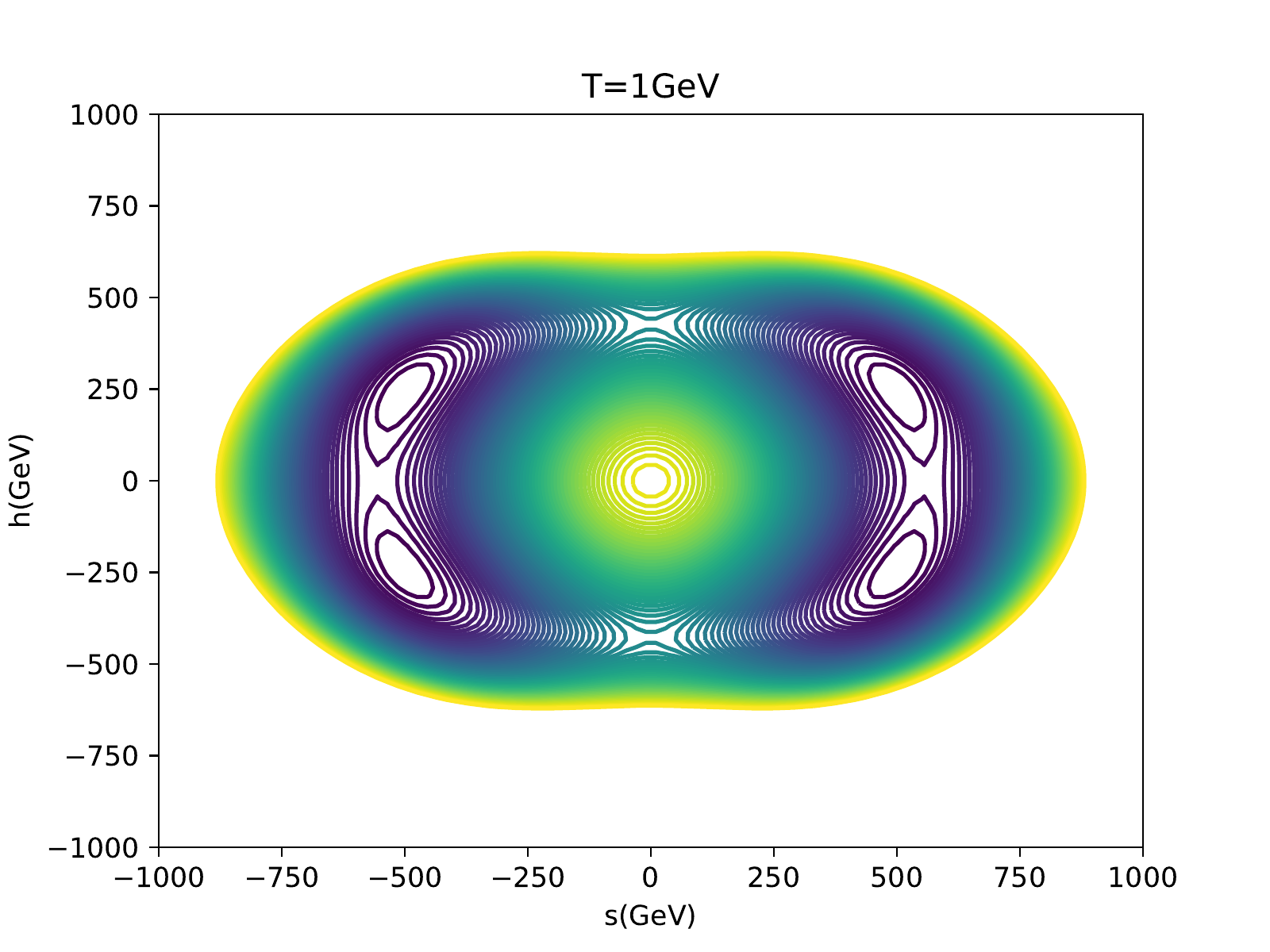}
}
\caption{An example of a $s$-direction first-order PT (notice the zoomed-in plots) followed by a higher-order PT along the mixing direction. This is given by the parameters $v_{BL}=500~ {\rm GeV}$, $\lambda_s=0.2$, $\lambda_h=0.19$, $\lambda_{sh}=0.2028$, $g=1.75$. Here the $s$-direction PT happens at the temperature $T\simeq 357.08~{\rm GeV}$, the higher-order PT happens around $T\simeq 162~{\rm GeV}$.}
\label{S-first}
\end{figure*}

\begin{figure*}
\resizebox{0.33\textwidth}{!}{%
  \includegraphics{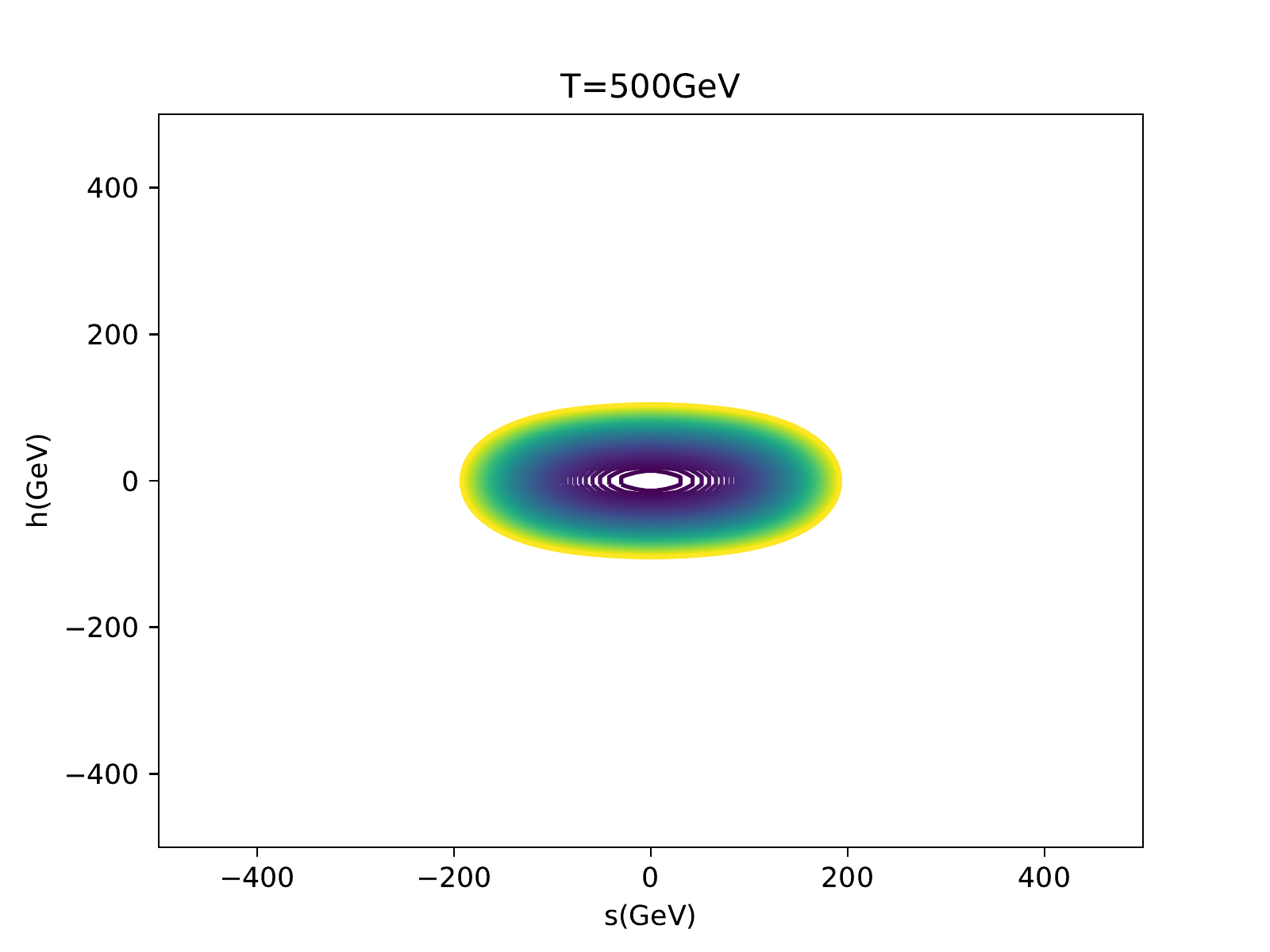}
}
\resizebox{0.33\textwidth}{!}{%
  \includegraphics{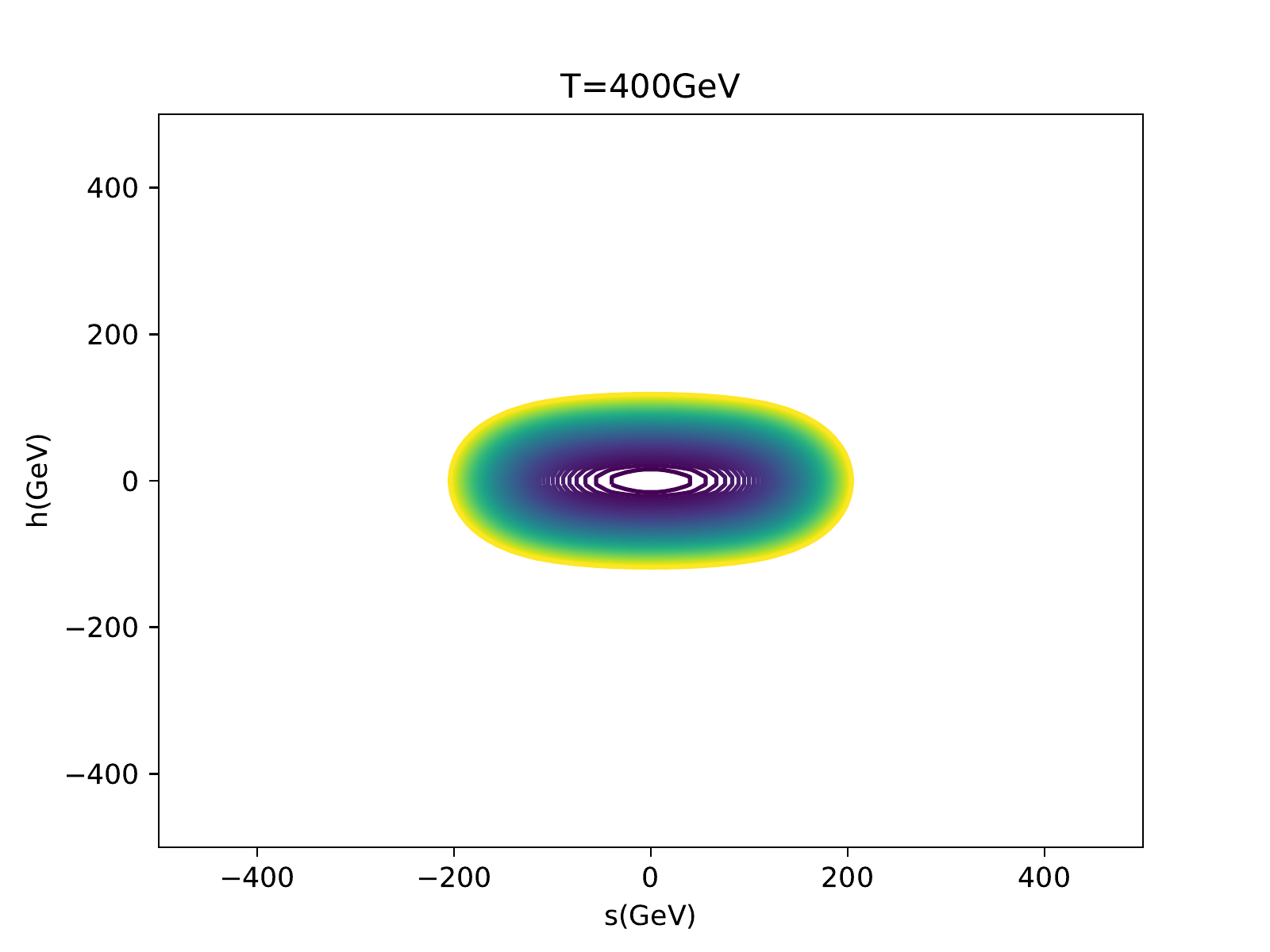}
}
\resizebox{0.33\textwidth}{!}{%
  \includegraphics{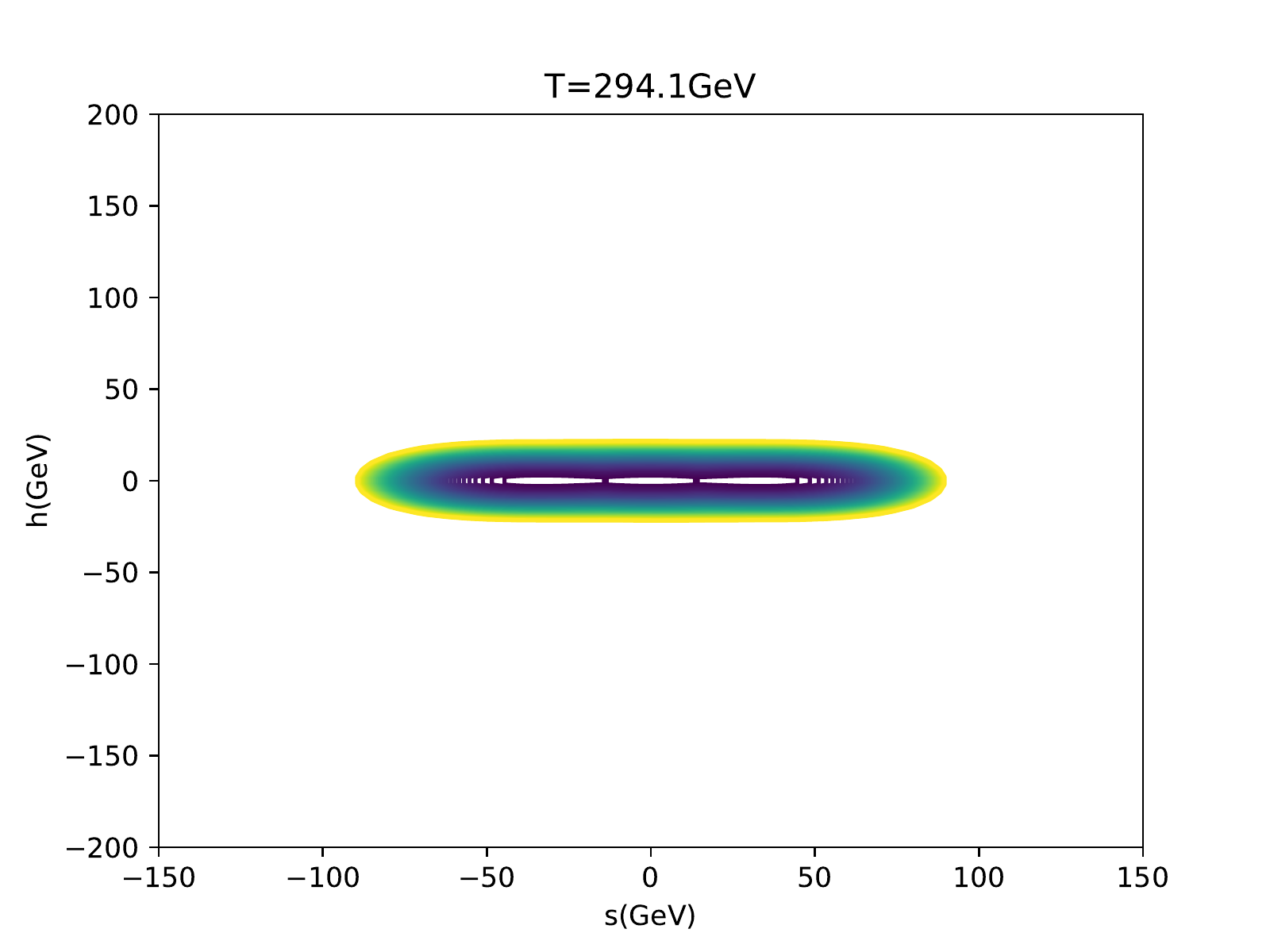}
}
\resizebox{0.33\textwidth}{!}{%
  \includegraphics{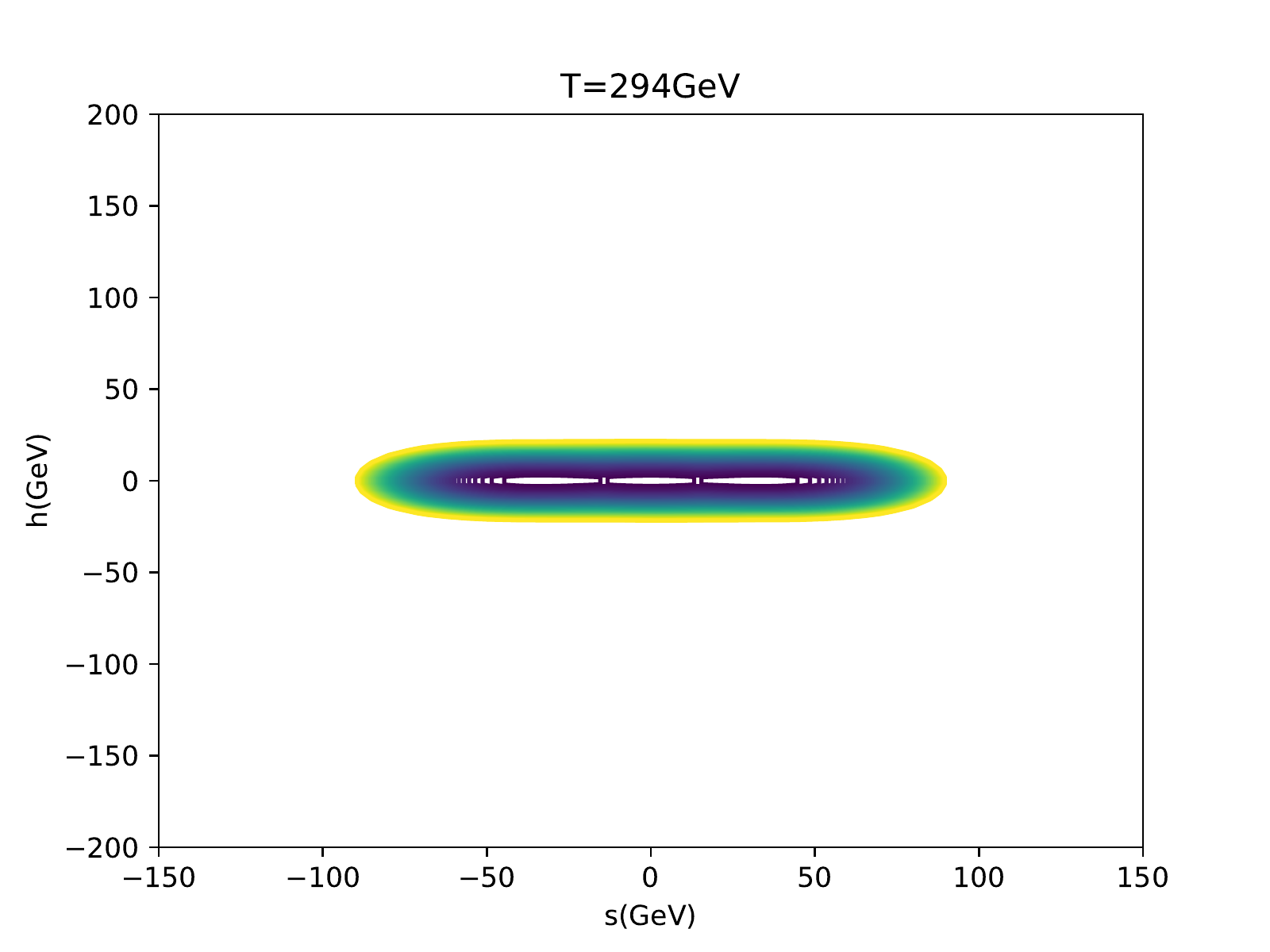}
}
\resizebox{0.33\textwidth}{!}{%
  \includegraphics{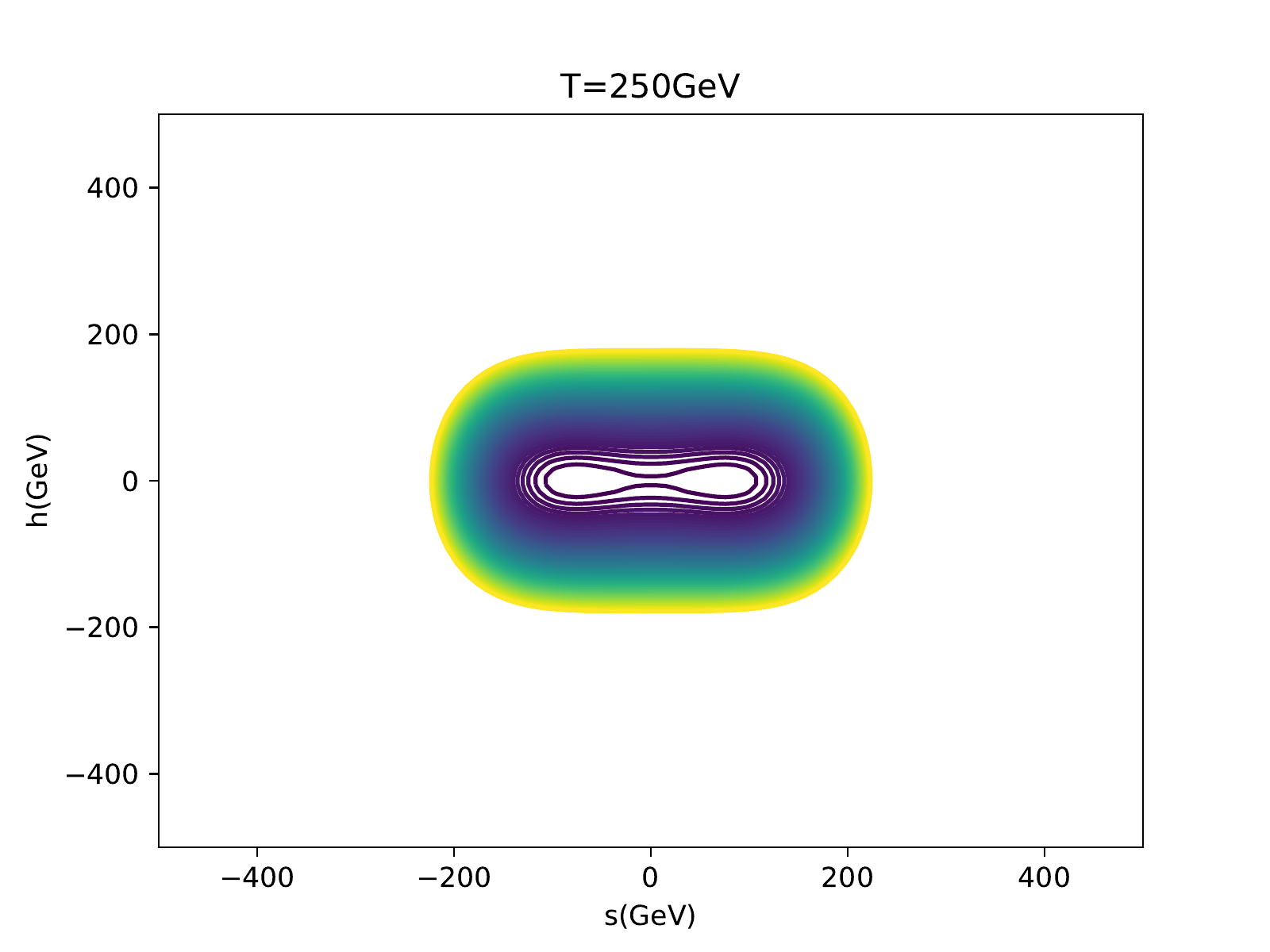}
}
\resizebox{0.33\textwidth}{!}{%
  \includegraphics{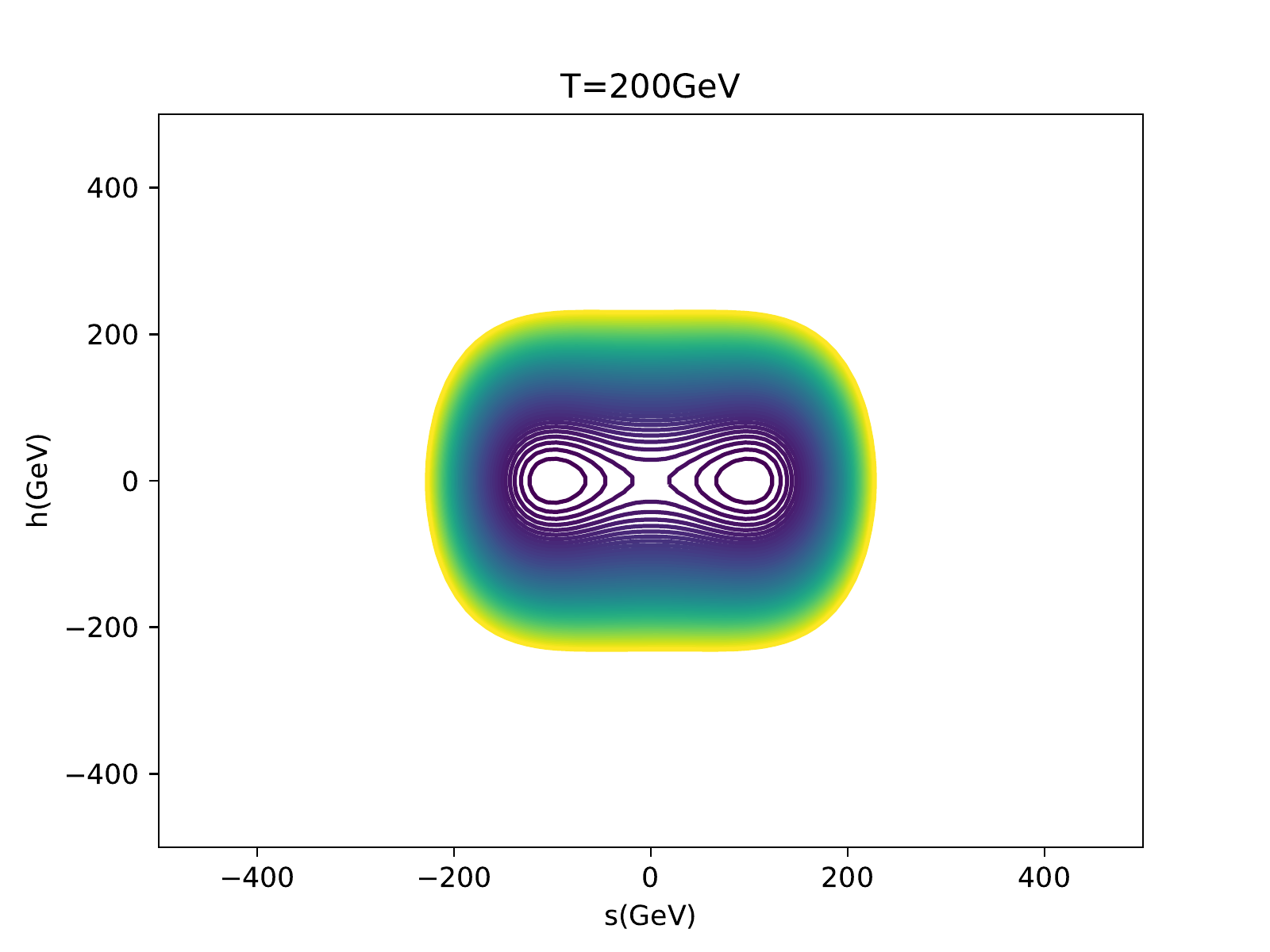}
}
\resizebox{0.33\textwidth}{!}{%
  \includegraphics{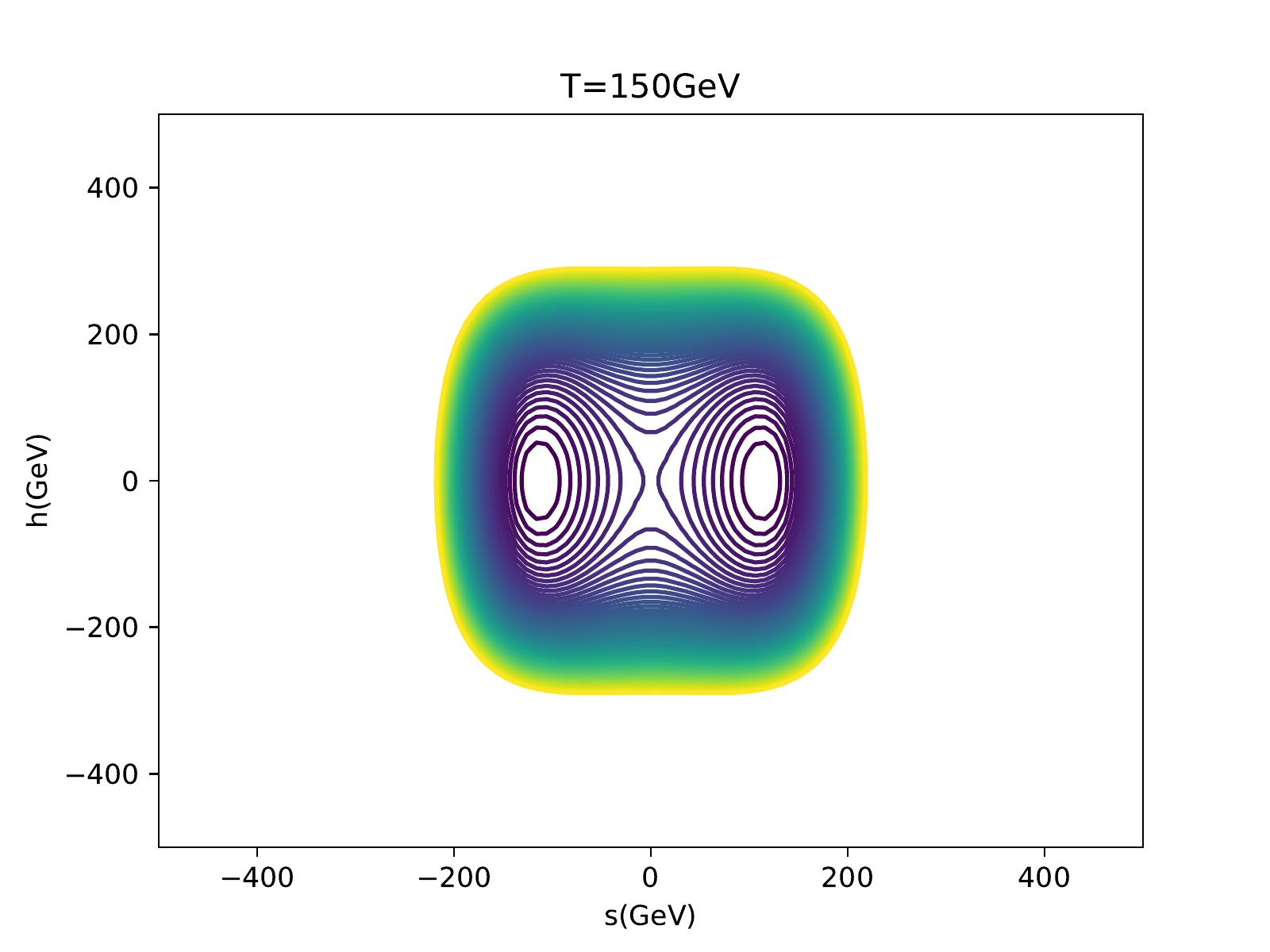}
}
\resizebox{0.33\textwidth}{!}{%
  \includegraphics{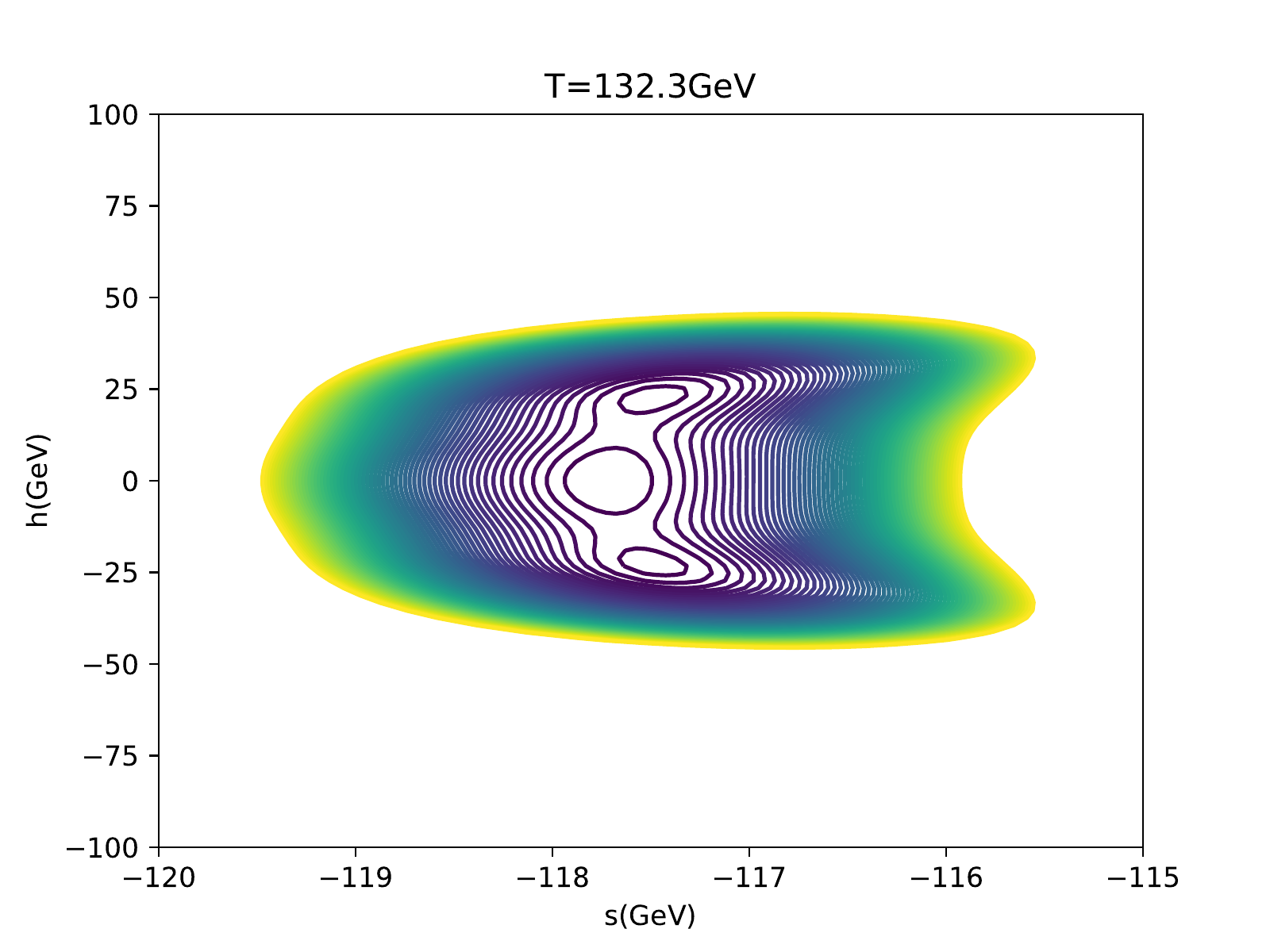}
}
\resizebox{0.33\textwidth}{!}{%
  \includegraphics{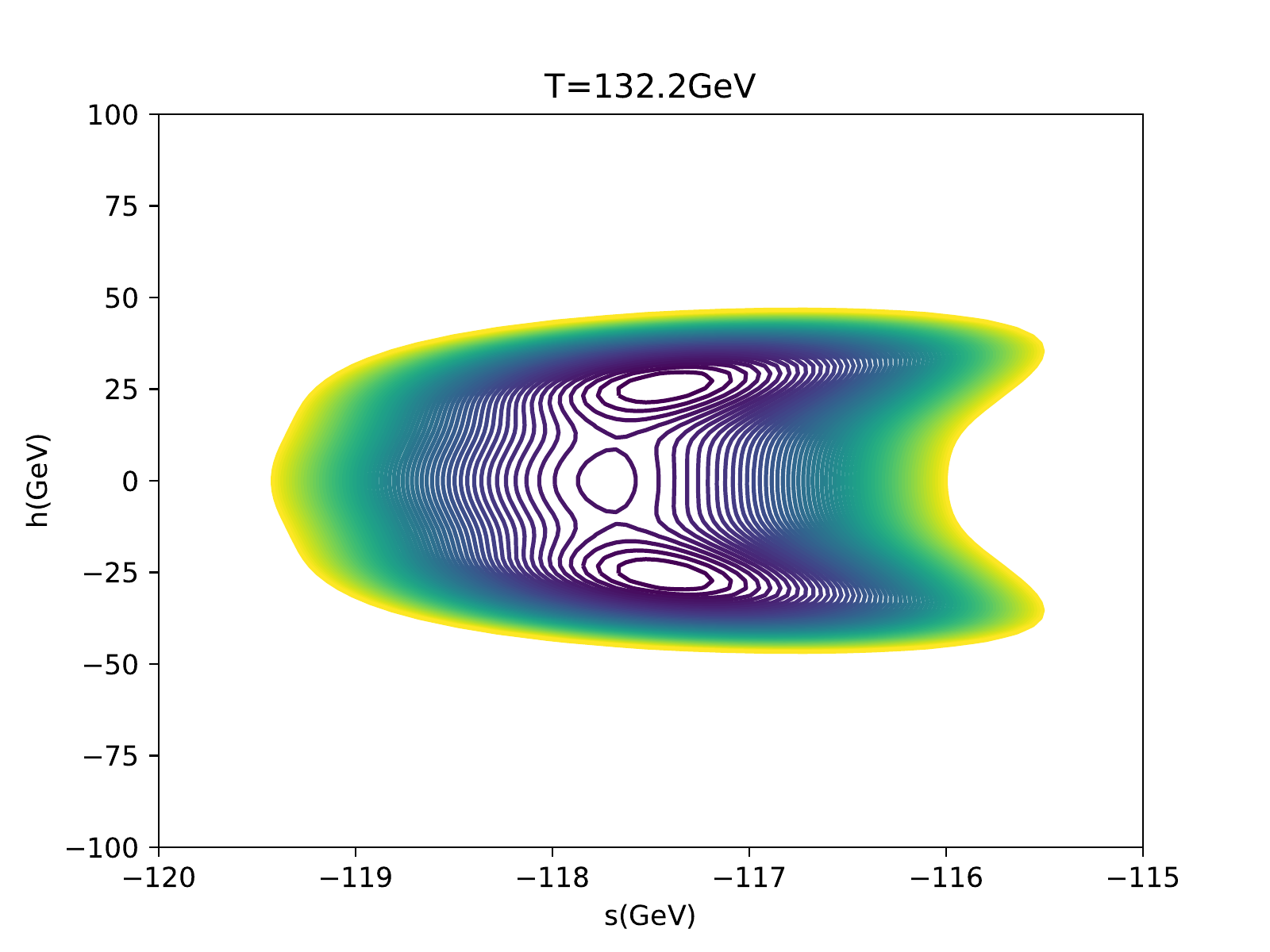}
}
\resizebox{0.33\textwidth}{!}{%
  \includegraphics{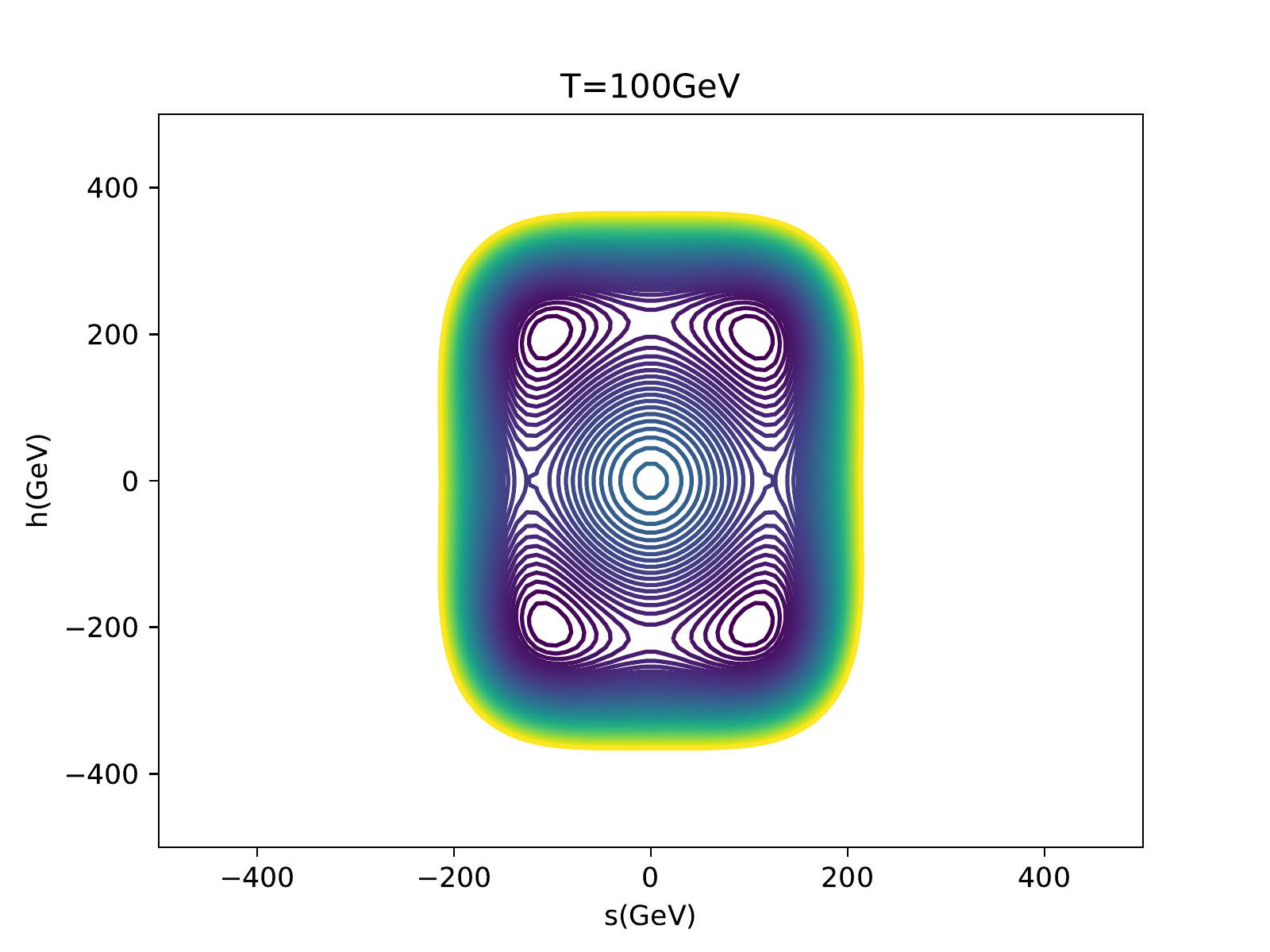}
}
\resizebox{0.33\textwidth}{!}{%
  \includegraphics{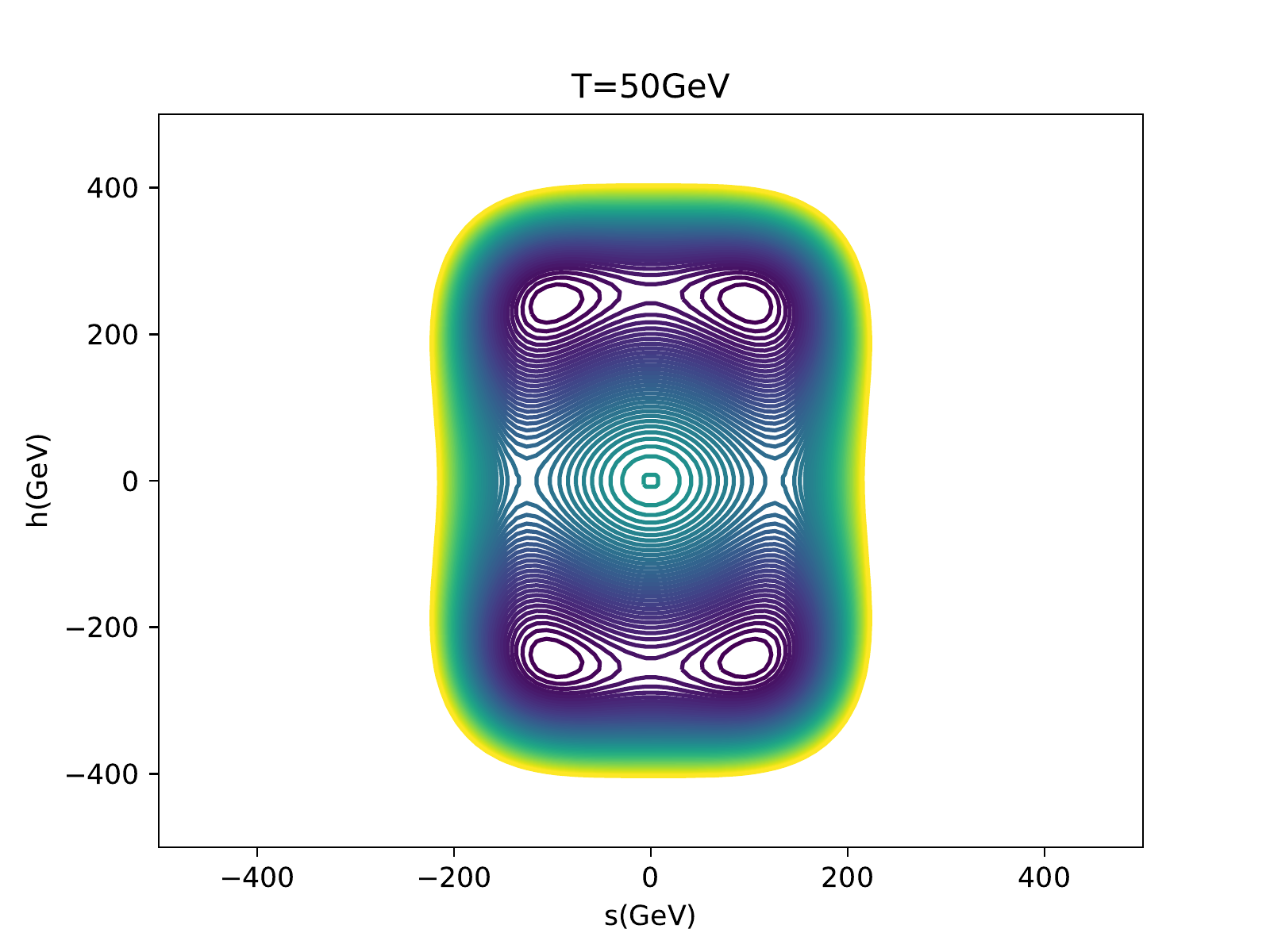}
}
\resizebox{0.33\textwidth}{!}{%
  \includegraphics{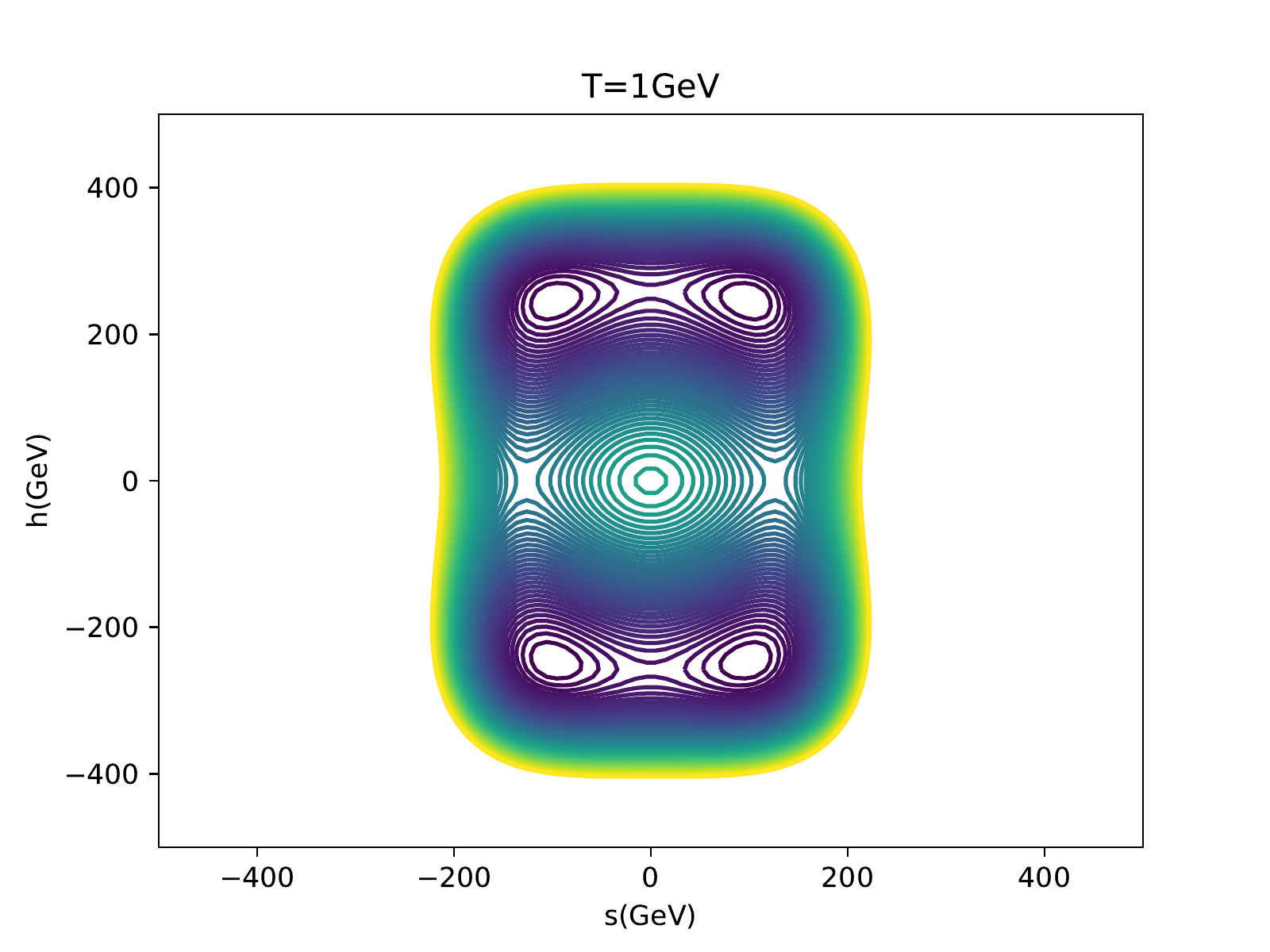}
}
\caption{An example of two first-order PTs, the first is along the $s$-direction followed by the second which is along the mixing direction (notice the zoomed-in plots). This is given by the parameters $v_{BL}=100~{\rm GeV}$, $\lambda_s=0.5$, $\lambda_h=0.12$, $\lambda_{sh}=0.1012$, $g=0.2$. Here, the $s$-direction PT happens at temperature $T\simeq 294.06~{\rm GeV}$, the hybrid PT along the mixing direction happens at $T\simeq 132.23~{\rm GeV}$.}
\label{SH}
\end{figure*}

In the singlet Majoron model, parameters can be divided into two distinct groups: the SM ones and the beyond-SM ones. For the SM parameters, we take the gauge couplings $g_1=0.65$ and $g_2=0.35$, the top quark's Yukawa coupling constant $y_t=0.989$, today's electroweak vacuum expected value (VEV) $v_{ew}=246~{\rm GeV}$; the only underdetermined parameter is the doublet Higgs boson's self-coupling constant $\lambda_{h}$, which may be shifted away from $m^2_{H, {\rm SM}}/(2\times v^2_{ew})$ due to the mixing effect. The beyond-SM parameters are the newly introduced singlet Higgs self-coupling $\lambda_{s}$, the mixing coupling $\lambda_{sh}$, the Dirac mass term Yukawa coupling $f$, the Majorana mass term Yukawa coupling $g$, and the VEV of the global symmetry $v_{BL}$. It easy to see that the parameter $f$ is degenerate with $g$ and $v_{BL}$ once a light neutrino mass scale is imposed; we thus will not set it as a free parameter. Finally, we have five parameters to be constrained: $v_{BL},~\lambda_{s},~\lambda_{h},~\lambda_{sh},~g$. For convenience in plotting, we will use a numeric string such as `0.3-0.124-0.24-1' to represent $\lambda_s=0.3, \lambda_h=0.124, \lambda_{sh}=0.24, g=1$, for $v_{BL}$, we will show it by setting the title of the subplot in Figs. \ref{GW01} and \ref{GW02}.

For direct comparison of theoretical and experimental constraints in Refs. \cite{Clarke:2013aya, Falkowski:2015iwa, Robens:2015gla}, we can remap $( v_{BL}, \lambda_s, \lambda_h, \lambda_{sh})$ into $( \tan\beta, m_1^2, m^2_2, \sin\theta )$ according to the formulas in Appendix \ref{appA}---notice only three of them are independent after $m_1^2$ or $m^2_2$ takes the SM Higgs boson's mass. As a matter of fact, we find the most convenient path is as follows: firstly give values to $v_{BL}$, $\lambda_s$ and $\lambda_h$, then use their relationship to evaluate $\lambda_{sh}$. When the SM Higgs is the lighter eigenstate, i.e., $m_1 = 125 ~{\rm GeV}$, from Eq. (\ref{m1_m2}) we have
\begin{eqnarray}
\label{mass_SM_Higgs_be_lighter}
 m^2_2 = 2\left(\lambda_s v^2_{BL} +\lambda_h v^2_{ew}\right) -m^2_1 ~,
\end{eqnarray}
and in the opposite case, $m_2 = 125 ~{\rm GeV}$, we can use
\begin{eqnarray}
\label{mass_SM_Higgs_be_heavier}
 m^2_1 = 2\left(\lambda_s v^2_{BL} +\lambda_h v^2_{ew}\right) -m^2_2 ~,
\end{eqnarray}
to evaluate $m_1$. In both cases
\begin{eqnarray}
\label{lambda_sh}
 \lambda_{sh} = \left[ \frac{1}{4} \left( \frac{m^2_2 -m^2_1}{v_{ew} v_{BL}} \right)^2 -\left( \lambda_s \frac{v_{BL}}{v_{ew}} -\lambda_h \frac{v_{ew}}{v_{BL}} \right)^2 \right]^{1/2} ~.
\end{eqnarray}

In this research we scan the following parameter space:
\begin{eqnarray}
\label{ParameterSpace}
 25 ~{\rm GeV} \leq v_{BL} \leq 1000 ~{\rm GeV} ~,~~ \\ \nonumber
 0 \leq \lambda_s ~,~ \lambda_h~, ~\lambda_{sh} \leq 1 ~,~~ 0 \leq g \leq 2 ~,
\end{eqnarray}
and by doing so we directly compare the parameters with the results in Ref. \cite{Robens:2015gla}. In Ref. \cite{Robens:2015gla}, the authors separate the parameter space into three regions: the high mass (HM) region with $m_2 \in \left[130,1000\right]{\rm GeV}$, the intermediate mass (IM) region with $m_{1,2} \in \left[120,130\right]{\rm GeV}$, and the low mass (LM) region with $m_1 \in \left[1,120\right]{\rm GeV}$. According to their constraints (see Figs. 10, 16, and 18 in Ref. \cite{Robens:2015gla}), we are permitted to take an upper bound for the U(1) symmetry breaking scale: $\tan\beta \lesssim 10$, or $25~{\rm GeV}\lesssim v_{BL}$. In order to cover all the parameter space, we will employ the following scan strategy.
\begin{itemize}
\item{\bf Step I}: start from the minima of $v_{BL}$, $\lambda_s$, and $\lambda_h$.
\item{\bf Step II}: evaluate the mass of the (non)SM Higgs from Eq. (\ref{mass_SM_Higgs_be_lighter}) or Eq. (\ref{mass_SM_Higgs_be_heavier}), $\lambda_{sh}$ from Eq. (\ref{lambda_sh}), $\sin\theta$ and $\tan\beta$ from Appendix \ref{appA}.
\item{\bf Step III}: compare $( \tan\beta, m_1^2, m^2_2, \sin\theta )$ with the theoretical and experimental constraints. If they can pass, then go to {\bf Step IV}, otherwise iterate $v_{BL}$, $\lambda_s$ and $\lambda_h$ and go to {\bf Step II}.
\item {\bf Step IV}: input the parameters $(v_{BL}, \lambda_{s}, \lambda_{h}, \lambda_{sh}, g)$ ($g$ starts form its initial value) into the model and calculate the PTs, iterate $g$,
\item{\bf Step V}: iterate $v_{BL}$, $\lambda_s$ and $\lambda_h$, initialize $g$, go to {\bf Step II}, or end when all the parameter points have been exhausted.
\end{itemize}

Since there are two scalar fields, the effective potential of Eq. (\ref{effective potential}) is a 2D surface and evolves with changing temperature. In the early universe the temperature is high, symmetries are unbroken, the zero point of the fields' space turns out to be the only vacuum. As the universe expands its temperature falls, the structure of the effective potential of Eq. (\ref{effective potential}) becomes nontrivial and different local minima will be formed, among them the lowest one is the true vacuum. PTs can happen by tunneling from the higher vacua to lower ones. Sometimes there are barriers between the distinct vacua, which correspond to first-order PTs, but at some other times PTs can happen very smoothly without any potential barriers being formed; these are referred as higher-order PTs. In our numerical simulation, we find that patterns of PTs in the singlet Majoron model can be quite fruitful, usually there are multi-step PTs, some of them are along the $s$ field direction, some are along the $h$ field direction, and some others are along a mixing direction. Since we are mostly interested in the strong first-order PTs, we will show some examples in the figures.

In Fig. \ref{Hybrid} a higher-order PT occurs around $T\simeq 182~{\rm GeV}$, along the $s$ field direction, followed by a strong first-order PT at the temperature $T\simeq 118.89~{\rm GeV}$. We see that as temperature falls, the original vacuum located at the origin point splits into two symmetric new vacua on the Higgs axis; this is a higher-order PT. When the temperature drops even further, two new vacua on the h axis are formed. These coexist with the old vacua until the tunneling occurs at $T\simeq 118.89~{\rm GeV}$. This first-order PT is very strong; here it tunnels from one vacuum located on the first axis directly to the second one located on another axis. Due to the mixing effect between Higgs and the $s$ field, hereafter we shall refer to these as 'hybrid PTs'. In our simulation, we do find some parameters which can lead to very strong hybrid PTs and detectable GWs for the mentioned interferometers. Unfortunately, they also lead to unacceptable branch ratio for the $H\rightarrow hh$ decay channel. For the parameter in Fig. \ref{Hybrid}, it predicts $m_1 = 7.9~{\rm GeV}$, and ${\rm BR}_{H\rightarrow hh}\simeq 0.99$, which definitely has been ruled out by collider experiments. We present it here in order to completely show the 2D PT properties of the singlet Majoron model.

Fig. \ref{S-first} shows a first-order PT along the $s$-direction (notice the zoomed-in plots) followed by a higher-order PT along the mixing direction.
The occurrence of the $s$-direction PT can also be attributed to the mixing coupling $\lambda_{sh}$, from Fig. \ref{S-first} one can see $s/T \sim 100/357 < 1$, which means the PT is weak.
Indeed, it proceeds very fast, leads to very high frequency GWs which go beyond the detectable range of spatial GW interferometers. As mentioned before, the final vacuum is located at $(v_{\rm BL}, ~v_{\rm ew})$ obviously when the temperature gets close to 0. This parameter gives $m_2 \simeq 328 ~ {\rm GeV}$ and thus belongs to the high mass region (see the details in Table \ref{Table-III}).

It is even more interesting to find multiple first-order PTs from some parameters. In Fig. \ref{SH} we show an example with a hybrid first-order PT happening after a $s$-direction first-order PT. Near $T\simeq 294.06~{\rm GeV}$, there exist three vacua separated by barriers, one located at the zero point; the other two are the nonzero ones with opposite signs according to the symmetry. As temperature drops, the field tunnels from the origin to the nonzero VEV. Notice here we also employ zoomed-in plots to better show the two first PTs. As above the PTGWs have too high frequency and too weak amplitude to be detectable.  The other first-order PT happens to be a hybrid one at $T\simeq 132.23~{\rm GeV}$; its GWs are also too weak to be detectable. This point also belongs to the low mass region (see the details in Table \ref{Table-II}).

We also find that, for some parameter space, it would allow weak first-order PTs above the $U(1)$ symmetry breaking, but the perturbative analysis has become jeopardized if the Higgs VEV in the broken phase is too small. To address this issue, one probably needs to do a lattice analysis (namely following \cite{Kajantie:1995kf}) that would find a crossover in this parameter space. However, as we have seen from the above analyses, the patterns of cosmological PTs in a singlet Majoron model are already very fruitful, which is a distinct feature for 2D problems compared to single field ones. We thus believe that the single Majoron model is a useful laboratory to study the properties of 2D cosmological PTs and GWs. In the follow-up study we plan to develop our numerical method further as regards the lattice analysis on the issue of non-perturbative regime in order to explore more details on cosmological PTs and GWs in higher dimensions.

\section{Phase transition gravitational waves in the singlet Majoron model}
\label{PTGWs}

\begin{figure*}
\centering
\resizebox{0.475\textwidth}{!}{%
  \includegraphics{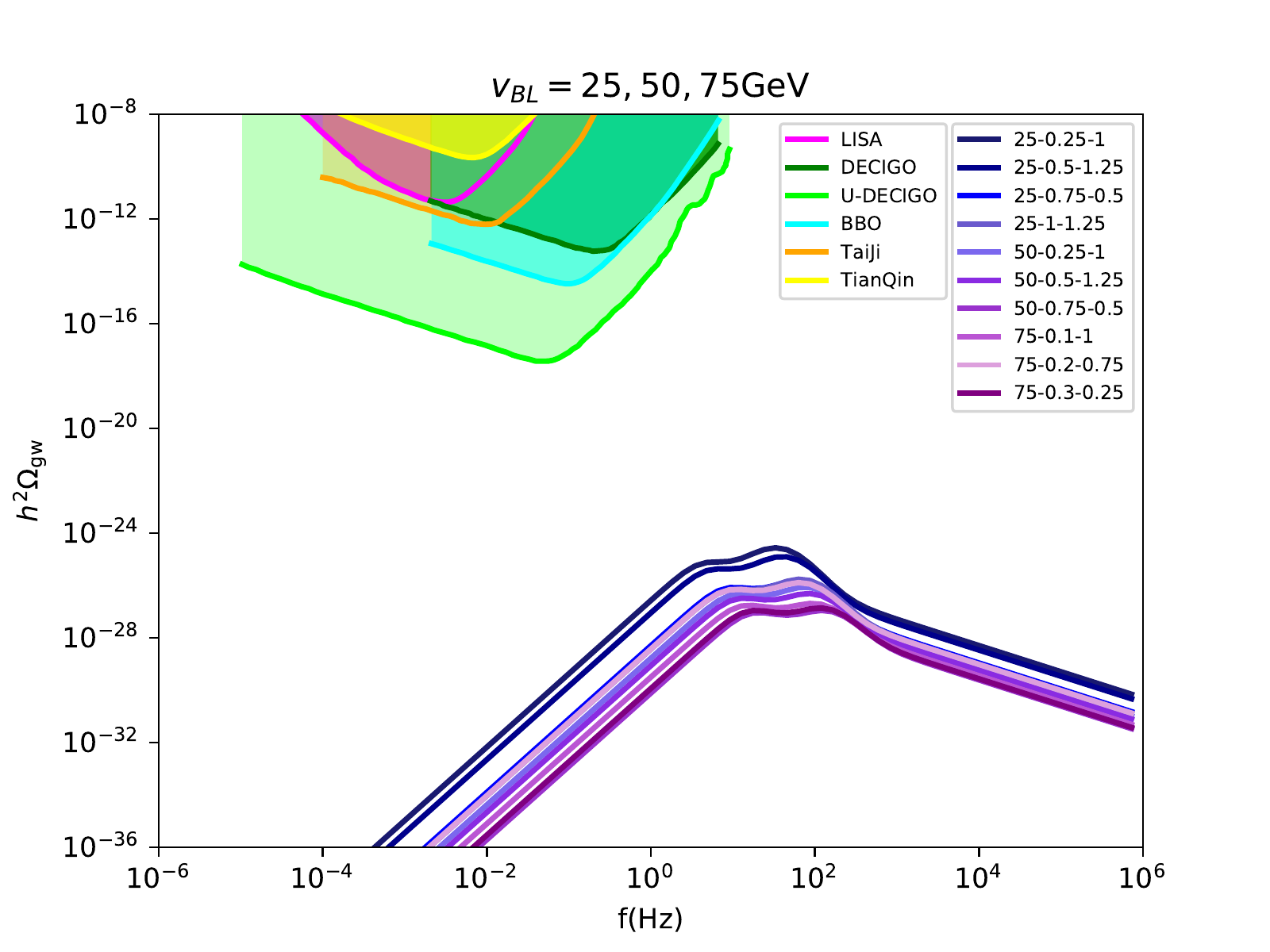}
}
\resizebox{0.475\textwidth}{!}{%
  \includegraphics{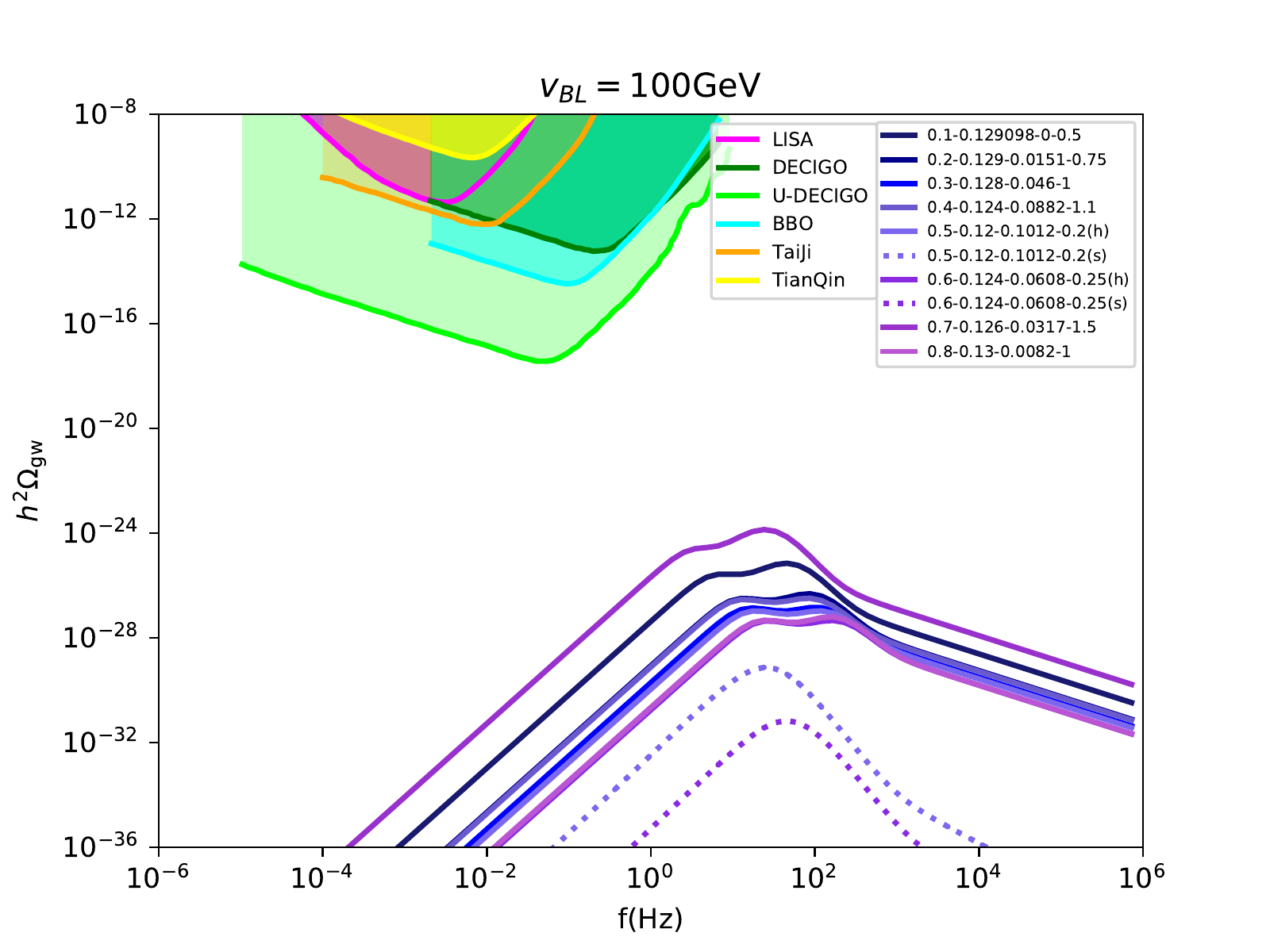}
}
\resizebox{0.475\textwidth}{!}{%
  \includegraphics{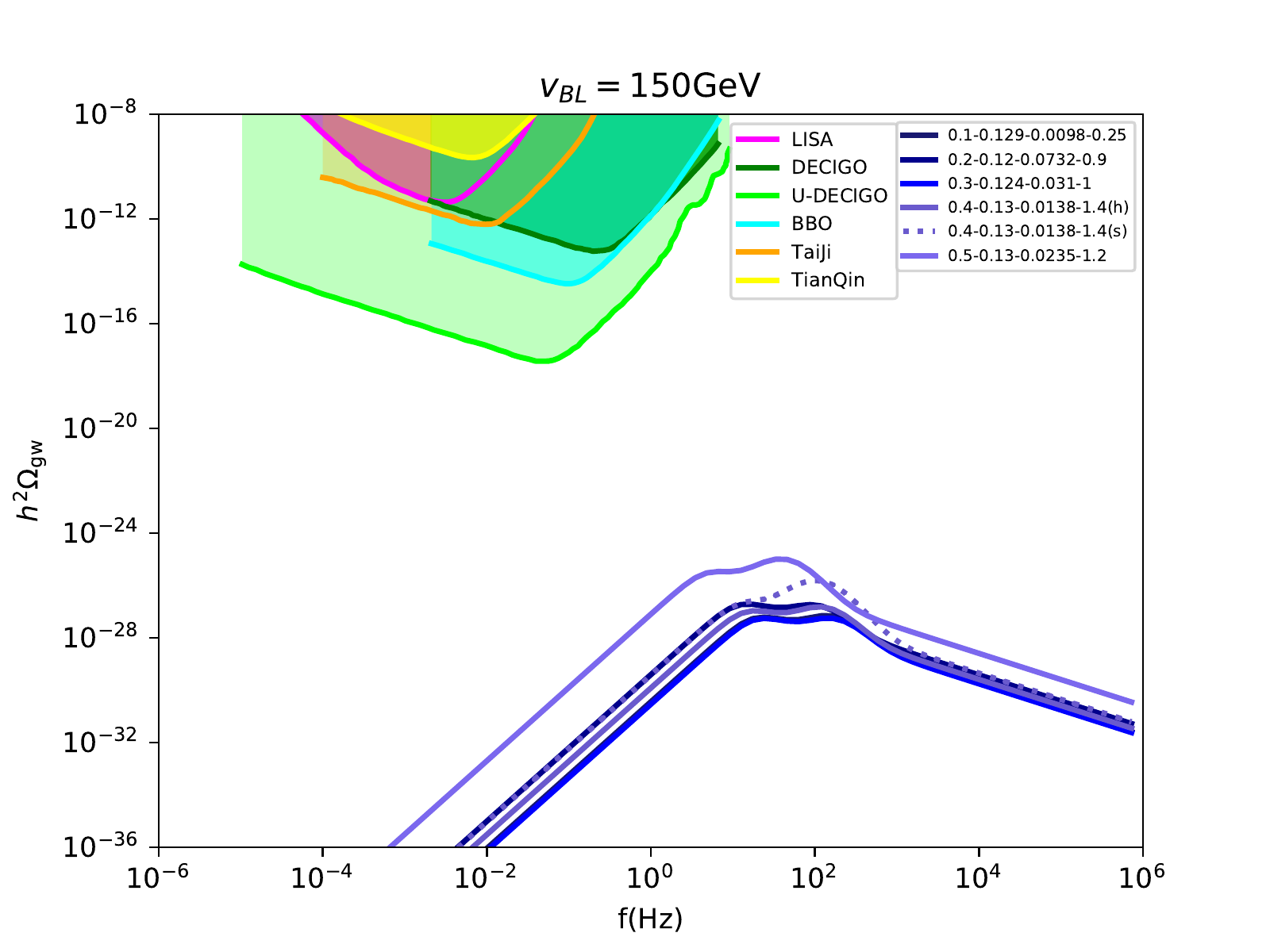}
}
\resizebox{0.475\textwidth}{!}{%
  \includegraphics{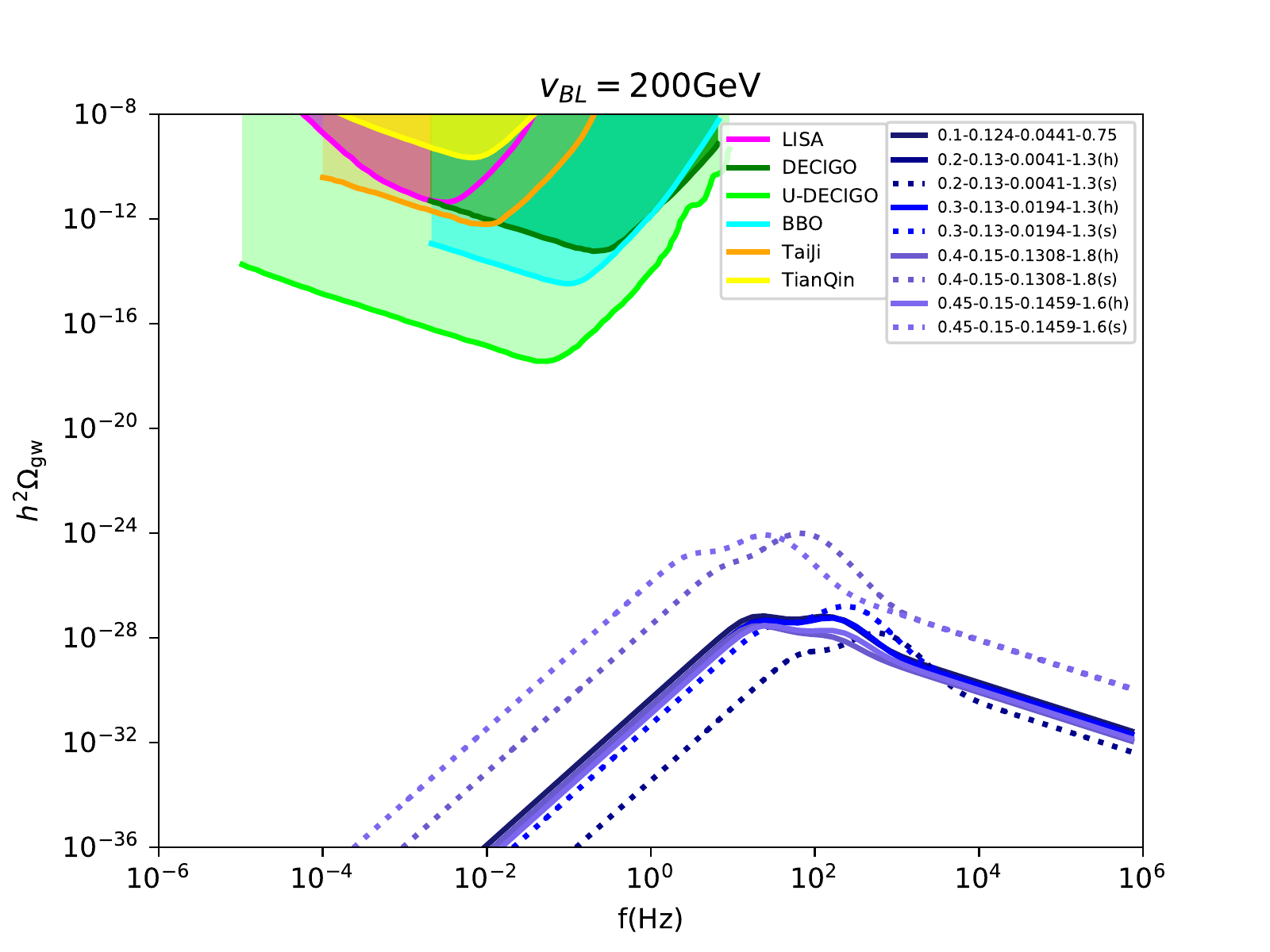}
}
\renewcommand{\figurename}{Fig.}
\caption{Gravitational wave signals from the model parameters with $v_{BL} <v_{ew} =246~\rm{GeV}$, compared with the detectability of future spatial interferometers. For the benchmark scales $v_{BL}=25~, 50~$, and $75~{\rm GeV}$, we use a simplified string notation '$v_{BL}-\lambda_s  - g$' since we have fixed $\lambda_h = m^2_{H, {\rm SM}}/(2\times v^2_{ew})$, and $\lambda_{sh} = 0$. For the benchmark scales $100~$, $150~$, and $200~{\rm GeV}$, we take another notation `$\lambda_s -\lambda_h - \lambda_{sh} - g$' in the plots. The solid lines refer to GWs from hybrid first-order PTs, the dotted lines refer to $s$-direction PTs. For parameters that lead to first-order PTs twice, we have used suffixes $(h)$ or $(s)$ to make it clearer. }
\label{GW01}
\end{figure*}

\begin{figure*}
\centering
\resizebox{0.475\textwidth}{!}{%
  \includegraphics{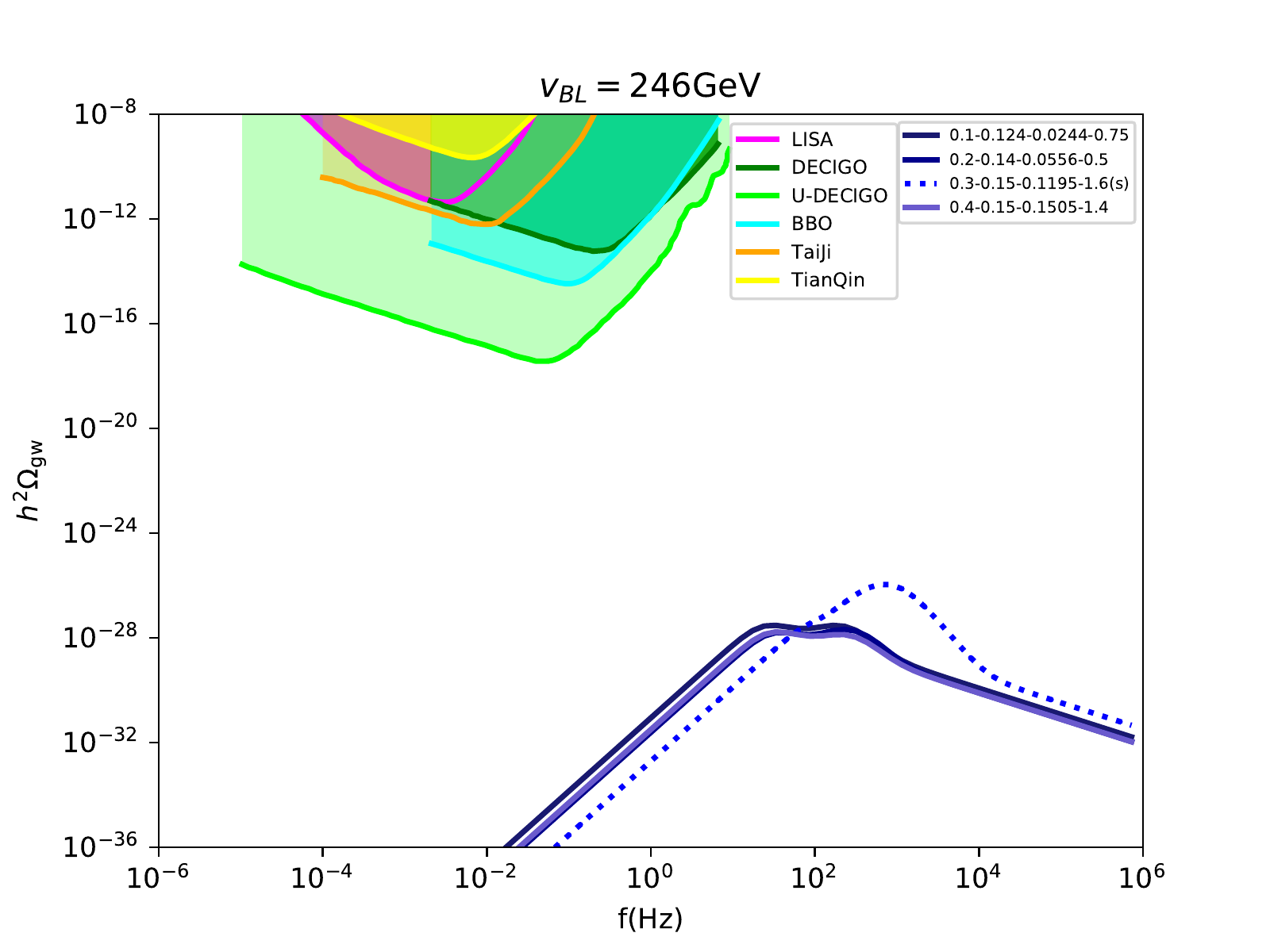}
}
\resizebox{0.475\textwidth}{!}{%
  \includegraphics{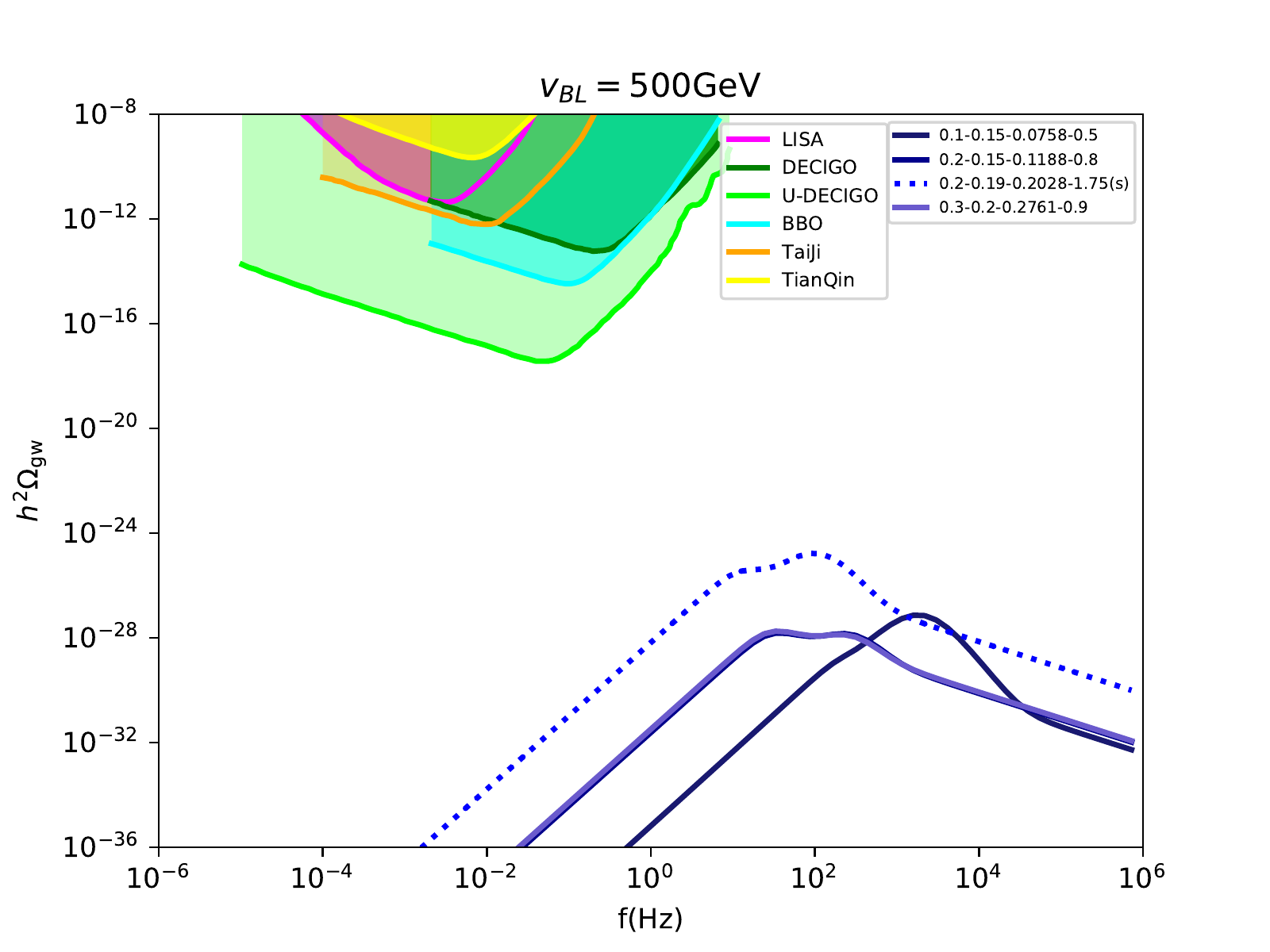}
}
\resizebox{0.475\textwidth}{!}{%
  \includegraphics{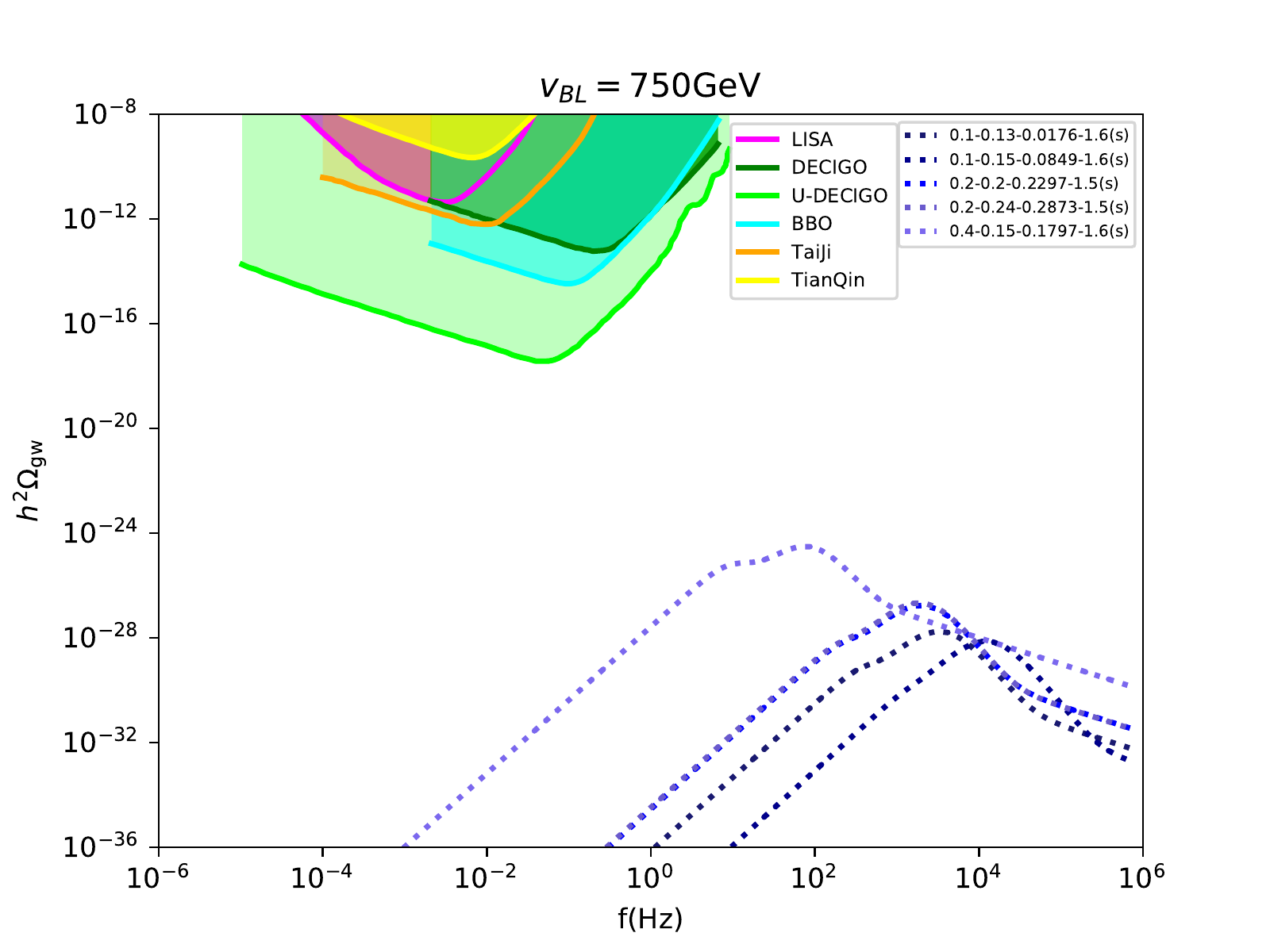}
}
\resizebox{0.475\textwidth}{!}{%
  \includegraphics{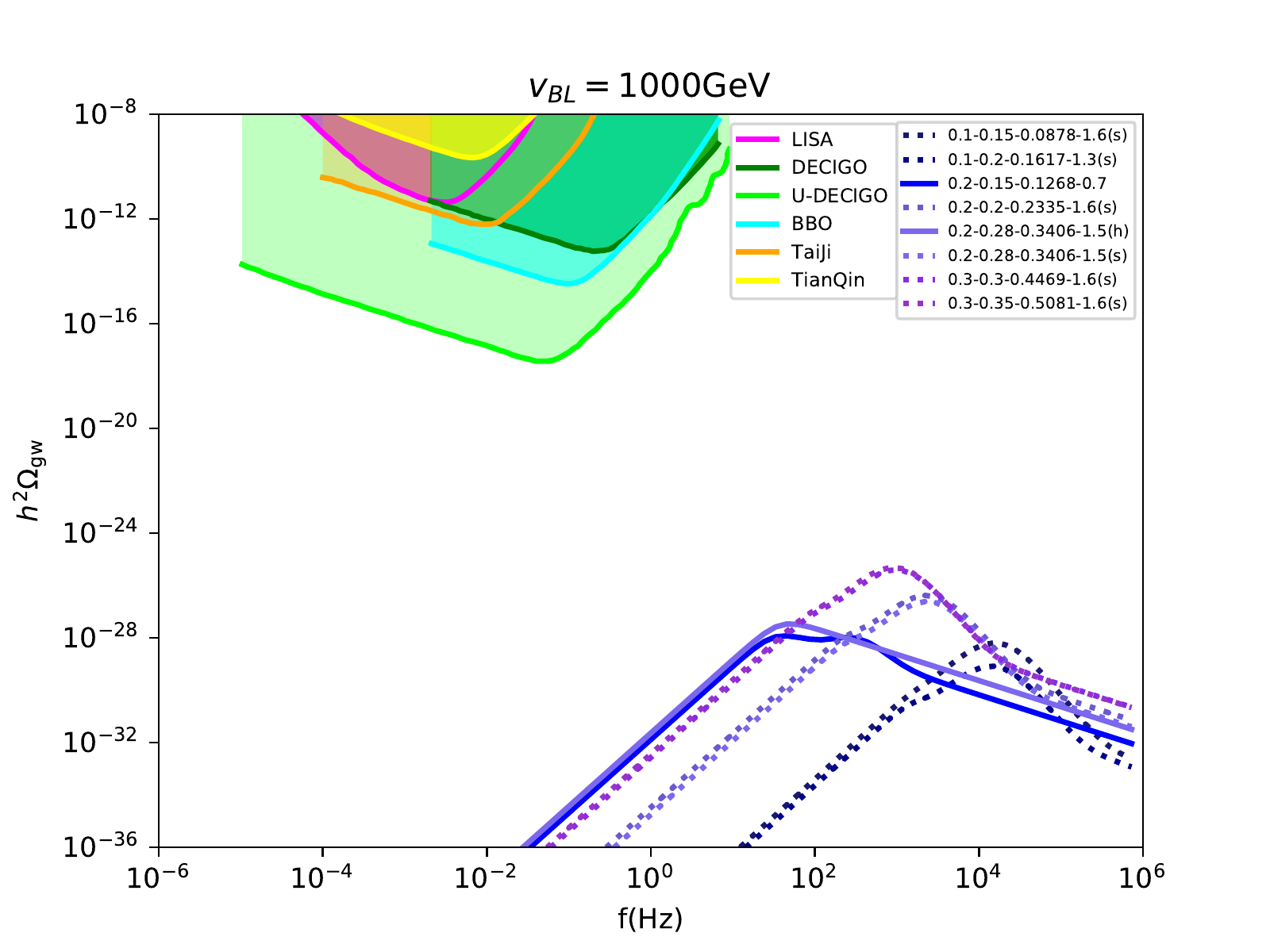}
}
\renewcommand{\figurename}{Fig.}
\caption{Gravitational wave signals from the model parameters with $v_{BL}\geq v_{ew}=246~\rm{GeV}$, compared with the detectability of future spatial interferometers. The benchmark scales are taken $v_{BL}=246~{\rm GeV}, 500~{\rm GeV}, 750~{\rm GeV},~1000~{\rm GeV}$, with the string notation `$\lambda_s-\lambda_h-\lambda_{sh}-g$'. The solid lines refer to GWs from hybrid PTs, the dotted lines refer to $s$-direction PTs. For parameters which lead to twice first-order PTs, we have used suffixes $(h)$ or $(s)$ to make it clearer.}
\label{GW02}
\end{figure*}

There are various GW sources generated during strong first-order PTs; we follow the study in \cite{Caprini:2015zlo} by considering the contributions from scalar fields, sound waves \cite{Hindmarsh:2013xza, Hindmarsh:2015qta, Hindmarsh:2017gnf}, and magnetohydrodynamic (MHD) turbulence \cite{Nicolis:2003tg}. The first GW source comes from the vacuum bubble collision \cite{Kamionkowski:1993fg, Nicolis:2003tg, Huber:2008hg}, the GW energy spectrum and peak frequency are
\begin{eqnarray}
 &\Omega_{\rm env} (f_{\rm env}) h^2 \simeq 1.67\times 10^{-5} \left( \frac{1}{\tilde{\beta}} \right)^{2} \left( \frac{\kappa_s \alpha}{1+\alpha} \right)^2 \left( \frac{100}{g_{*}} \right)^{\frac{1}{3}}  \\ \nonumber
 & \left( \frac{0.11v^3_w}{0.42+v^2_w} \right) \times \frac{3.8(f/f_{\rm env})^{2.8}}{1+2.8(f/f_{\rm env})^{3.8}} ~, \\
 &f_{\rm env} \simeq 1.65 \times 10^{-5} {\rm{Hz}} \left( \frac{0.62}{1.8-0.1v_w+v^2_w} \right) {\tilde{\beta}}  \left( \frac{T_*}{100~{\rm{GeV}}}\right) \\ \nonumber
 &\left( \frac{g_*}{100} \right)^{\frac{1}{6}} ~.
\end{eqnarray}
These formulas found by experience are summarized from numerical simulations with envelope approximation (also see the work with an analytic derivation \cite{Caprini:2007xq,Jinno:2016vai} or beyond the envelope approximation \cite{Cutting:2018tjt, Jinno:2017fby}). The second source is generated by the sound waves of the bulk motion \cite{Hindmarsh:2013xza, Hindmarsh:2015qta, Hindmarsh:2017gnf},
\begin{eqnarray}
\label{GWs from sound waves}
 &\Omega_{\rm sw} (f_{\rm sw}) h^2 \simeq 2.65 \times 10^{-6} \left( \frac{1}{\tilde{\beta}} \right) v_w \left( \frac{\kappa_{\nu} \alpha}{1+\alpha} \right)^2 \left( \frac{100}{g_{*}} \right)^{\frac{1}{3}} \\ \nonumber
 & \Big[ \frac{7(f/f_{\rm sw})^{\frac{6}{7}}}{4+3(f/f_{\rm sw})^2} \Big]^{\frac{7}{2}} ~, \\
\label{peak frequency_sound waves}
 &f_{\rm sw} \simeq 1.9\times 10^{-5} {\rm{Hz}} \left( \frac{1}{v_w} \right) {\tilde{\beta}} \left( \frac{T_*}{100~{\rm{GeV}}} \right)\left( \frac{g_*}{100} \right)^{\frac{1}{6}} ~.
\end{eqnarray}
The third are GWs generated by MHD turbulence \cite{Nicolis:2003tg} with energy spectrum and peak frequency
\begin{eqnarray}
 &\Omega_{\rm tu} (f_{\rm tu}) h^2 \simeq 3.35 \times 10^{-4} \left( \frac{1}{\tilde{\beta}} \right) v_w \left( \frac{\kappa_{\rm tu} \alpha}{1+\alpha}\right)^{\frac{3}{2}} \left( \frac{100}{g_{*}} \right)^{\frac{1}{3}}  \\ \nonumber
 &\frac{(f/f_{\rm tu})^3}{[1+(f/f_{\rm tu})]^{\frac{11}{3}} (1+8\pi f/h_{*})} ~, \\
 &f_{\rm tu} \simeq 2.7 \times 10^{-5} {\rm{Hz}} \left( \frac{1}{v_w} \right) {\tilde{\beta}} \left( \frac{T_*}{100~{\rm{GeV}}} \right) \left( \frac{g_*}{100} \right)^{\frac{1}{6}} ~,
\end{eqnarray}
where $h_* = 1.65 \times 10^{-5}{\rm Hz} \left(T_{*}/100~{\rm GeV} \right) \left(g_{*}/100 \right)^{\frac{1}{6}}$.

In Ref. \cite{Caprini:2015zlo} the authors discuss three different kinds of bubbles, i.e., the non-runaway bubbles, the bubbles with runaway in plasma, and the bubbles with runaway in vacuum. In this paper we will follow them directly (also see Ref. \cite {Espinosa:2010hh} for more details). For non-runaway bubbles, we evaluate the total GW energy spectrum from $\Omega_{\rm sw} (f_{\rm sw}) h^2+\Omega_{\rm tu} (f_{\rm tu}) h^2$, take the efficiency factor $\kappa_{\nu} = \alpha (0.73 + 0.083 \sqrt{\alpha}+\alpha)^{-1}$ when the bubble wall velocity is quite close to 1, or $\kappa_{\nu} = v^{\frac{6}{5}}_{w} 6.9 \alpha (1.36-0.037\sqrt{\alpha}+\alpha)^{-1}$ when $v_w$ is smaller than $0.1$; we take the turbulence efficiency factor $\kappa_{\rm tu} = 0.05 \kappa_{\nu}$. For the case of runaway bubbles in plasma, the contributions from the $s$ field itself is also included, and the smallest $\alpha$ in this case can be found by calculating
\begin{eqnarray}
 \alpha_{\infty} \simeq \frac{30}{24\pi^2} \frac{\sum_a c_a \Delta m^2_a (s_*, h_*)}{g_{{\rm eff}*}T^2_{*}} ~,
\end{eqnarray}
where the squared mass difference between the two phases can be read in Appendix \ref{appA}. The coefficients $c_a$ satisfy \\
$(c_{H},~c_{G},~c_{\rho},~c_{\chi},~c_{W},~c_{Z},~c_{t},~c_{\nu_R}) =(1,~3,~1,~1,~6,~3,~6,~3)$. The efficiency factor is evaluated from $\kappa_{\nu}=(\alpha_{\infty}/\alpha)\kappa_{\infty}$, with $\kappa_{\infty}=\alpha_{\infty} /\left( 0.73 +0.083 \sqrt{\alpha_{\infty}} +\alpha_{\infty}\right)$. When $\phi_*/T_*$ is very large, the PT may not have happened until the universe is super-cooled, which is quite different from the non-runaway case or the case of runaway in plasma. In our paper, we will not deal with runaway in vacuum bubbles, since the criterion of $T_{*}$ of Eq. (\ref{criterion of Tstar}) is spoiled when the total energy density of the universe is dominated by the vacuum energy.

Carrying out our scan strategy, for parameters with the same $(v_{BL},~\lambda_s)$, we pick out the ones which can produce the largest GW amplitudes and display their signals in Fig. \ref{GW01} with $v_{BL} < 246~{\rm GeV}$ and Fig. \ref{GW02} with $v_{BL} \geq 246~{\rm GeV}$, meanwhile confronting them with the space-borne interferometers such as LISA, DECIGO, BBO, TAIJI and TianQin. Note that there are too many curves in some subplots; in order to save the plots space we will take the unique string notation such as `$\lambda_s-\lambda_h-\lambda_{sh}-g$'. For the benchmark parameters taken in Figs. \ref{GW01} and \ref{GW02}, we also make three tables (Tables \ref{Table-I}, \ref{Table-II} for $v_{BL} < 246~{\rm GeV}$ and Table \ref{Table-III} for $v_{BL} \geq 246~{\rm GeV}$) in Appendix \ref{appB} to show more details, especially as regards which mass region(HM, IM and LM) they belong to. We can use the three tables to analyze how they could survive from the theoretical and experimental constraints such as in Ref. \cite{Robens:2015gla}.

When $v_{BL} < 100~{\rm GeV}$, the most of the benchmark parameters lead to (non-)SM Higgs with mass $m_1 < m_{H, {\rm SM}}/2$, and they thus are severely constrained by the collider experiments on the decay channel $H\rightarrow hh$. Since the upper bound about the decay branch ratio ${\rm BR}_{H\rightarrow hh} \lesssim 0.2$ \cite{Aad:2015txa, Khachatryan:2016whc},  the coupling $\lambda_h$ must be quite close to the SM value $m^2_{H, {\rm SM}}/(2\times v^2_{ew})$, and $\sin\theta$ nearly equals $1$. In order to survive from the $H\rightarrow hh$ constraints, in Table \ref{Table-I} we take three benchmark scales, $v_{BL}=25~{\rm GeV}$, $50~{\rm GeV}$, and $75~{\rm GeV}$, set $\lambda_h=m^2_{H, {\rm SM}}/(2\times v^2_{ew})$, and meanwhile we let the mixing coupling $\lambda_{sh}$ vanish. From Eq. (\ref{mass_SM_Higgs_be_heavier}) we can get an upper bound \\
$\lambda_s \le (1/4) \lambda_{h,{\rm SM}}(v_{ew}/v_{BL})^2$, which equals $3.1$, $0.78$ and $0.35$ for $v_{BL}=25~{\rm GeV}$, $50~{\rm GeV}$ and $75~{\rm GeV}$. In the second quadrant of Fig. \ref{GW01}, we show the largest GWs generated for the benchmark scales. All the display curves are due to the hybrid first-order PTs. Unluckily these GWs are hardly within reach of the planned interferometers.

For $100~{\rm GeV}\leq v_{BL} < 246~{\rm GeV}$, all the three cases (LM, IM and HM) can be realized. In Table \ref{Table-II} we show the parameters which have been taken in Fig. \ref{GW01} for benchmark scales $v_{BL} = 100{\rm GeV}, ~150{\rm GeV}, ~200{\rm GeV}$. Taking advantage of Fig. 7 in Ref. \cite{Robens:2015gla}, one can see that a large value (close to unity) for $\sin \theta$ is needed for the LM cases, while a small value (smaller than $\sim 0.4$) is needed for the HM cases. We have compared the parameters in Table \ref{Table-II} to make sure that they are not ruled out by the experiments. Notice again that there exists an upper bound for $\lambda_s$, i.e., $\lesssim 0.8$ for $100{\rm GeV}$, $\lesssim 0.5$ for $150{\rm GeV}$, and $\lesssim 0.45$ for $200{\rm GeV}$. This is due to the constraint contour on $\tan\beta$ and $m_2$ in the HM region; see Fig. 10(b) in Ref. \cite{Robens:2015gla}. Compared with the smaller $v_{BL}$ in Table \ref{Table-I}, there tend to be more and more GWs generated by multiple first-order PTs.

In Fig. \ref{GW02} we show the GWs generated by parameters with  $v_{BL} \geq 246~{\rm GeV}$; see Table \ref{Table-III} for more details as regards the chosen parameters. All except one belong to the HM region, the upper bound for $\lambda_s$ becomes smaller and drops slower. As we have mentioned, for these high $v_{BL}$ parameters, the PTs along the $s$-direction play a more and more important role in generating GWs, with even higher peak frequencies.

\section{Conclusion}
\label{conclusion}

In this paper, we study the properties of cosmological PTs, especially the resulting GWs, in the singlet Majoron model, using a numerical simulation to treat the two-field problem without freezing any of the field directions. Compared with a single-field treatment, the patterns of PTs turn out to be much more diverse. We have not only verified the pattern with an EWPT happening after the global $U(1)$ symmetry breaking, but also we find new patterns, such as strong hybrid PTs happening before the $U(1)$ symmetry breaking. Our simulation suggests that the single Majoron model is an ideal benchmark in understanding the phenomenology of two-field cosmological PTs.

The PTGWs are likely not within the reach of detectability of space-borne interferometers such as LISA, DECIGO, BBO, TAIJI and TianQin---either their amplitudes are too low or their frequencies are too high for those next-generation instruments. As a matter of fact, without considering the collider experimental exclusion bounds, we are able to find strong hybrid PTs which can generate detectable GWs. Unfortunately, in such cases, the mixing coupling necessarily has to be large, and the non-SM Higgs need acquire a mass smaller than the half mass of the heavier Higgs. Experimentally this parameter space is ruled out by constraints on the $H\rightarrow hh$ decay branch ratio. However, we emphasize that the aforementioned conclusion is model dependent and the numerical method developed in the present study shall be widely applied to another model of cosmological PTs, which involves more than one degrees of freedom.

Finally, we would like to highlight the implications of the developed numerical method, which could be extended from several perspectives in future study. Note that we have only considered the bubbles which belong to the types of non-runaway or runaway in plasma (see Ref. \cite{Caprini:2015zlo}), but we abandon those bubbles with runaway in vacuum (see Ref. \cite{Bodeker:2009qy}; see also Ref. \cite{Ellis:2018mja} for a most recent study), since they spoil our criterion in Eq. (\ref{criterion of Tstar}). By doing so we expect that the parameter points would be further constrained. Additionally, the same analysis may be applied to some cosmological PTs that occur at extremely low frequency band and hence might be falsifiable in the forthcoming experiments of primordial gravitational waves \cite{Cai:2016hqj, Li:2017drr, Cai:2007xr, Cai:2014uka}.

\begin{acknowledgements}
We are grateful to Andrea Addazi, Jim Cline, Ryusuke Jinno, Antonino Marciano and Pierre Zhang for valuable comments.
We also thank two reviewers for insightful suggestions on the manuscript.
YPW would like to thank Carroll L. Wainwright for useful discussions of the package of CosmoTransitions.
This work is supported in part by the NSFC (nos. 11722327, 11653002, 11421303, J1310021), by CAST Young Elite Scientists Sponsorship Program (2016QNRC001), by the National Youth Thousand Talents Program of China, by the Fundamental Research Funds for the Central Universities, and by the Postdoctoral Science Foundation of China (2017M621999).
All numerical simulations are operated on the computer clusters Linda \& Judy in the particle cosmology group at USTC.
\end{acknowledgements}

\section{Mass spectrum, cut-off, and Debye mass}
\label{appA}

The field-dependent mass spectrum can be found as follows:
\begin{eqnarray}
\label{mass spectrum}
& m^2_{\rho\rho}(s,h) = \lambda_s \left(3s^2-v_{BL}^2\right)+\frac{1}{2}\lambda_{sh}\left(h^2-v_{ew}^2\right);\\
& m^2_{\chi\chi}(s,h) = \lambda_s \left(s^2-v_{BL}^2\right)+\frac{1}{2}\lambda_{sh}\left(h^2-v_{ew}^2\right);\\
& m^2_{HH}(s,h) = \lambda_h \left(3h^2-v_{ew}^2\right)+\frac{1}{2}\lambda_{sh}\left(s^2-v_{BL}^2\right);\\
& m^2_{GG}(s,h) = \lambda_h \left(h^2-v_{ew}^2\right)+\frac{1}{2}\lambda_{sh}\left(s^2-v_{BL}^2\right);\\
& m^2_{\nu_R}(s,h) = \frac{1}{2}g^2 s^2;~~~m^2_{t}(s,h)=\frac{1}{2}y_t^2 h^2;\\
& m^2_{W}(s,h) = \frac{1}{4}g_1^2 h^2;~~~m^2_{Z}(s,h)=\frac{1}{4}\left(g_1^2+g_2^2\right) h^2;\\
& m^2_{\rho H}(s,h) = \lambda_{sh}sh;
\end{eqnarray}
where the subscripts `$H$' and `$G$' represent for the Higgs boson and Goldstones of the Higgs doublet. One can see that $m^2_{\chi\chi}=0$ when $(s,~h) = (v_{BL},~v_{ew})$, and thus in the current vacuum the Goldstone of $\sigma$ is massless when there are no higher dimensional effective terms.

Since H and $\rho$ are mixed, we can introduce the mixing angle by
\begin{eqnarray}
\left(
  \begin{array}{c}
    h_1 \\
    h_2 \\
  \end{array}
\right)
=
\left(
  \begin{array}{cc}
    \cos\theta & -\sin\theta \\
    \sin\theta  & \cos\theta\\
  \end{array}
\right)
\left(
  \begin{array}{c}
    H \\
    \rho \\
  \end{array}
\right) ~.
\end{eqnarray}
After diagonalization, the lighter and heavier new Higgs bosons get masses as follows:
\begin{eqnarray}
&m^2_{1,2}=  \\ \nonumber
&\frac{1}{2}(m^2_{\rho\rho}+m^2_{HH})\mp\frac{1}{2}\sqrt{(m^2_{\rho\rho}-m^2_{HH})^2+4(m^2_{\rho H})^2} ~,
\end{eqnarray}
respectively. Today we have $(s,~h)=(v_{BL},~v_{ew})$, and hence,
\begin{eqnarray}
\label{m1_m2}
&m^2_{1(0)} = \\ \nonumber
&\left(\lambda_s v^2_{BL} +\lambda_h v^2_{ew}\right)
-\sqrt{\left(\lambda_s v^2_{BL} -\lambda_h v^2_{ew}\right)^2 +\lambda_{sh}^2 v^2_{ew}v^2_{BL}} ~, \\
&m^2_{2(0)} =  \\ \nonumber
&\left(\lambda_s v^2_{BL} +\lambda_h v^2_{ew}\right)
+\sqrt{\left(\lambda_s v^2_{BL} -\lambda_h v^2_{ew}\right)^2 +\lambda_{sh}^2 v^2_{ew}v^2_{BL}} ~.
\end{eqnarray}
Here the subscript `${(0)}$' refers to the results in the present vacuum. We can discard it
without bringing about any ambiguity, and then today's mixing angle satisfies
\begin{eqnarray}
&\sin 2\theta = \frac{\lambda_{sh}v_{ew}v_{BL}}
{\sqrt{\left(\lambda_s v^2_{BL}-\lambda_h v^2_{ew}\right)^2+\lambda_{sh}^2 v^2_{ew}v^2_{BL}}} ~, \\
&\cos 2\theta = \frac{\lambda_{s}v^2_{BL}-\lambda_{h}v^2_{ew}}
{\sqrt{\left(\lambda_s v^2_{BL}-\lambda_h v^2_{ew}\right)^2+\lambda_{sh}^2 v^2_{ew}v^2_{BL}}} ~.
\end{eqnarray}
In order to compare with other work, let us also introduce another angle,
\begin{eqnarray}
\tan \beta \equiv \frac{v_{ew}}{v_{BL}} ~.
\end{eqnarray}

Consider a counter term
\begin{eqnarray}
\Delta V_{ct}^{\rm{T=0}}(s,h)=A h^2 ~,
\end{eqnarray}
from the renormalization conditions of Eq. (\ref{RenormalizationConditions}), we can get the energy scale cut-off and coefficient $A$,
\begin{eqnarray}
&\log Q^2 = \left(\sum_i n_i m^2_i \frac{\partial m^2_i}{\partial s}\right)^{-1} \\ \nonumber
&\left[ \sum_i n_i m^2_i \frac{\partial m^2_i}{\partial s} \left( \log m^2_i -c_i +\frac{1}{2} \right) \right]
\Bigg{|}_{(v_{BL}, v_{ew})} \\
&A = -\frac{1}{64\pi^2 h}  \\ \nonumber
&\left[ \sum_i n_i m^2_i \frac{\partial m^2_i}{\partial h}
\left( \log m^2_i -\log Q^2 -c_i +\frac{1}{2} \right) \right]\Bigg{|}_{ (v_{BL}, v_{ew})}
\end{eqnarray}
where $\left(c_i, c_{W},c_{Z}\right)=\left(3/2,~5/6,~5/6\right)$.

The self-energies are given by
\begin{eqnarray}
&&\Pi_{\rho}(T) = \left( \frac{1}{3}\lambda_s +\frac{1}{6}\lambda_{sh} +\frac{1}{8}g^2 \right) T^2 ~, \\
&&\Pi_{\chi}(T)= \Pi_{\rho}(T) ~, \\
&&\Pi_{H}(T) = \left[ \frac{1}{16} ( 3g_1^2 +g^2_2) + \frac{1}{2} \lambda_h +\frac{1}{4}y^2_t +\frac{1}{12}\lambda_{sh} \right] T^2,\\
&&\Pi_{G}(T) = \Pi_{H}(T), \\
&&\Pi_{W_L}(T) = \frac{11}{6}g^2_1 T^2 ~,~~ \\
&&\Pi_{W_T}(T) = \Pi_{Z_T}(T) = \Pi_{\gamma_T}(T) = 0 ~,
\end{eqnarray}
and, the Debye mass for the longitudinal component of the $Z$ boson is
\begin{eqnarray}
&{\cal M}^2_{Z_L}(s,h,T) = \\ \nonumber
&\frac{1}{2} \left[ m^2_{Z}(s,h) +\frac{11}{6}\frac{g_1^2}{\cos^2{\theta_w}}T^2
+\Delta(s,h,T) \right] ~,\\
&\Delta(s,h,T) =  \\ \nonumber
& \left[ m^4_Z(s,h) +\frac{11}{3}\frac{g^2_1 \cos^2{2\theta_w}}{\cos^2{\theta_w}} \left( m^2_Z(s,h)
+\frac{11}{12} \frac{g_1^2}{\cos^2{\theta_w}} T^2 \right) T^2 \right]^{\frac{1}{2}} ~.
\end{eqnarray}

\section{The benchmark parameters}
\label{appB}

In the second part of the appendix, we show the benchmark parameters in Figs. \ref{GW01} and \ref{GW02}, which are presented in Tables \ref{Table-I}, \ref{Table-II}, and \ref{Table-III}. Apart from the model parameters, we also provide the results for the parameterization of $\left(\tan\beta, m_1^2, m^2_2, \sin\theta\right)$, to directly compare them with theoretical and experimental constraints such as in Ref.\cite{Robens:2015gla}. Following \cite{Robens:2015gla}, we have classified the parameter samples to be three groups, which are the LM, IM, and HM regions, respectively.

\begin{table*}
\centering
\caption{\label{Table-I} Part parameters in Fig. \ref{GW01}.  Here we have set
$\lambda_h = m^2_{H, SM}/(2\times v^2_{ew})$, and $\lambda_{sh} = 0$. LM: low mass region; IM: intermediate mass region; HM: high mass region.}
\begin{tabular}{|c|c|c|c|c|c|c|c|c|c|}
\hline
\hline
$v_{BL}~({\rm GeV})$     &$\lambda_{s}$     & $\lambda_{h}$     &$\lambda_{sh}$     &$g$
& $m_1~({\rm GeV})$     & $m_2~({\rm GeV})$     & $\tan {\beta}$     & $\sin{\theta} $   &{\rm Type}\\
\hline
25     & 0.25     & 0.129098     & 0.0     &1.0        & 17.68     & 125     &9.84     &1.0     &{\rm LM}   \\
\hline
25     & 0.5     & 0.129098     & 0.0     &1.25       & 25.0     & 125     &9.84     &1.0     &{\rm LM}   \\
\hline
25     & 0.75     & 0.129098     & 0.0      &0.5       & 30.62     & 125     &9.84     &1.0     &{\rm LM}   \\
\hline
25     & 1.0       & 0.129098     & 0.0       &1.25      & 35.36     & 125     &9.84     &1.0     &{\rm LM}   \\
\hline
50     & 0.25     & 0.129098     & 0.0       &1.0       & 35.36     & 125     &4.92     &1.0     &{\rm LM}   \\
\hline
50     & 0.5       & 0.129098     & 0.0       &1.25       & 50.0     & 125     &4.92     &1.0     &{\rm LM}   \\
\hline
50     & 0.75     & 0.129098     & 0.0       &0.5        &61.24     & 125     &4.92     &1.0      &{\rm LM}  \\
\hline
75     & 0.1       & 0.129098     & 0.0        &1.0        &33.54     & 125     &3.28     &1.0     &{\rm LM}   \\
\hline
75     & 0.2       & 0.129098     & 0.0        &0.75        &47.43     & 125     &3.28     &1.0     &{\rm LM}   \\
\hline
75     & 0.3       & 0.129098     & 0.0        &0.25        &58.09     & 125     &3.28     &1.0     &{\rm LM}   \\
\hline
\end{tabular}
\end{table*}

\begin{table*}
\centering
\caption{\label{Table-II}More parameters in Fig. \ref{GW01}.  LM: low mass region; IM: intermediate mass region; HM: high mass region.}
\begin{tabular}{|c|c|c|c|c|c|c|c|c|c|}
\hline
\hline
$v_{BL}~({\rm GeV})$     &$\lambda_{s}$     & $\lambda_{h}$     &$\lambda_{sh}$     &$g$
& $m_1~({\rm GeV})$     & $m_2~({\rm GeV})$     & $\tan {\beta}$     & $\sin{\theta} $   &{\rm Type}\\
\hline
100     & 0.1       & 0.129098     & 0.0        &0.5        &44.72     & 125     &2.46     &1.0     &{\rm LM}   \\
\hline
100     & 0.2     & 0.129     & 0.0151        &0.75      & 63.15      & 125    &2.46     &0.999490     &{\rm LM}   \\
\hline
100     & 0.3     & 0.128     & 0.0460       &1.0      & 76.60     & 125     &2.46     &0.993167     &{\rm LM}   \\
\hline
100     & 0.4     & 0.124     & 0.0882       &1.1       & 85.92     & 125     &2.46     &0.961840     &{\rm LM}   \\
\hline
100     & 0.5     & 0.120     & 0.1012       &0.2       & 94.33     & 125     &2.46     &0.914487     &{\rm LM}   \\
\hline
100     & 0.6     & 0.124     & 0.0608      &0.25       &106.69     & 125     &2.46     &0.924415      &{\rm LM}  \\
\hline
100     & 0.7     & 0.126     & 0.0317       &1.5       &116.73     & 125     &2.46     &0.901395     &{\rm LM}   \\
\hline
100     & 0.8     & 0.13      & 0.0082       &1.0          &125        & 126.92     &2.46     &0.474829     &{\rm IM}   \\
\hline
150     & 0.1     & 0.129     & 0.0098      &0.25       &66.99       &125        &1.64      &0.999467     &{\rm LM}   \\
\hline
150     & 0.2     & 0.12      & 0.0732       &0.9        &88.88     &125       &1.64     &0.926000     &{\rm LM}   \\
\hline
150     & 0.3     & 0.124     & 0.031       &1.0         &113.5     &125       &1.64     &0.880325     &{\rm LM}    \\
\hline
150     & 0.4     & 0.13     & 0.0138       &1.4         &125        &134.57    &1.64     &0.209624     &{\rm HM}    \\
\hline
150     & 0.5     & 0.13     & 0.0235       &1.2         &125        &150.36    &1.64     &0.125019     &{\rm HM}    \\
\hline
200     & 0.1      & 0.124    & 0.0441      &0.75       &85.92     &125          &1.23     &0.96184      &{\rm LM}   \\
\hline
200     & 0.2     & 0.13     & 0.0041       &1.3         &125        &126.92      &1.23     &0.474829     &{\rm IM}    \\
\hline
200     & 0.3     &0.13      &0.0194       &1.3         &125        &155.27      &1.23     &0.113430     &{\rm HM}    \\
\hline
200     & 0.4     & 0.15     & 0.1308       &1.8         &125        &185.82      &1.23     &0.365811     &{\rm HM}    \\
\hline
200     & 0.45     & 0.15     & 0.1459       &1.6         &125       &196.29      &1.23     &0.332338     &{\rm HM}     \\
\hline
\end{tabular}
\end{table*}

\newpage

\begin{table*}
\centering
\caption{\label{Table-III} Parameters in Fig. \ref{GW02}. LM: low mass region; IM: intermediate mass region; HM: high mass region.}
\begin{tabular}{|c|c|c|c|c|c|c|c|c|c|}
\hline
\hline
$v_{BL}~({\rm GeV})$     &$\lambda_{s}$     & $\lambda_{h}$     &$\lambda_{sh}$     &$g$
& $m_1~({\rm GeV})$     & $m_2~({\rm GeV})$     & $\tan {\beta}$     & $\sin{\theta} $   &{\rm Type}\\
\hline
246     & 0.1     & 0.124     & 0.0244   & 0.75     & 107.17     & 125     &1.0     &0.922451     &{\rm LM}\\
\hline
246     & 0.2     & 0.14     & 0.0556     & 0.5     & 125         & 159.77    &1.0     &0.365060     &{\rm HM}\\
\hline
246     & 0.3     & 0.15     & 0.1195      & 1.6     & 125         & 197.08    &1.0     &0.330114     &{\rm HM}\\
\hline
246     & 0.4     & 0.15     & 0.1505      & 1.4     & 125         & 225.70    &1.0     &0.267638     &{\rm HM}\\
\hline
500     & 0.1      & 0.15     & 0.0758     & 0.5     & 125         & 229.19     &0.49   &0.261819      &{\rm HM}\\
\hline
500     & 0.2      & 0.15     & 0.1188     & 0.8     & 125         & 320.20     &0.49   &0.170617      &{\rm HM}\\
\hline
500     & 0.2      & 0.19     & 0.2028     & 1.75     & 125         & 327.68     &0.49   &0.283447      &{\rm HM}\\
\hline
500     & 0.3      & 0.20     & 0.2761     & 0.9     & 125         & 398.22     &0.49   &0.245006      &{\rm HM}\\
\hline
750     & 0.1      & 0.13      & 0.0176     &1.6      & 125         & 335.57     &0.33   &0.033549      &{\rm HM}\\
\hline
750     & 0.1      & 0.15      & 0.0849     &1.6      & 125         & 339.16     &0.33   &0.159529      &{\rm HM}\\
\hline
750     & 0.2      & 0.2      & 0.2297     &1.5      & 125         & 483.30     &0.33   &0.198424      &{\rm HM}\\
\hline
750     & 0.2      & 0.24     & 0.2873     &1.5      & 125         & 488.29     &0.33   &0.245451      &{\rm HM}\\
\hline
750     & 0.4      & 0.15     & 0.1797     &1.6      & 125          & 672.70     &0.33   &0.076094      &{\rm HM}\\
\hline
1000    & 0.1      & 0.15     & 0.0878     &1.6      & 125          & 450.03     &0.25   &0.116341       &{\rm HM}\\
\hline
1000    & 0.1      & 0.2      & 0.1617      &1.3      & 125          & 456.71     &0.25   &0.210887        &{\rm HM}\\
\hline
1000    & 0.2      & 0.15      & 0.1268     &0.7      & 125         & 634.45     &0.25   &0.080861        &{\rm HM}\\
\hline
1000    & 0.2      & 0.2      & 0.2335     &1.6      & 125         & 639.20      &0.25   &0.147777        &{\rm HM}\\
\hline
1000    & 0.2      & 0.28      & 0.3406     &1.5      & 125         & 646.73      &0.25   &0.212980        &{\rm HM}\\
\hline
1000    & 0.3      & 0.3      & 0.4469     &1.6      & 125         & 787.84      &0.25   &0.184895        &{\rm HM}\\
\hline
1000    & 0.3      & 0.35     & 0.5081     &1.6      & 125         & 791.67      &0.25   &0.209165         &{\rm HM}\\
\hline
\end{tabular}
\end{table*}

\end{document}